\pdfoutput=1
\documentclass[%
 reprint,
 amsmath,amssymb,
prb,
]{revtex4-2}

\usepackage{graphicx}
\usepackage{dcolumn}
\usepackage{bm}

\usepackage{array, multirow}
\usepackage{xcolor}
\usepackage{txfonts}
\usepackage{hyperref}

\newcommand{\mat}[1]{\mathbf{#1}}

\newcommand{\tmps}{{TMPS$_3$}}
\newcommand{\mnps}{{MnPS$_3$}}
\newcommand{\feps}{{FePS$_3$}}
\newcommand{\nips}{{NiPS$_3$}}

\begin{document}


\title{Magnetic anisotropy and magnetic ordering of transition-metal phosphorus trisulfides}

\author{Tae Yun Kim}
\email{kimtaeyun@outlook.com}
\affiliation{Center for Correlated Electron Systems, Institute for Basic Science, Seoul 08826, Korea}
\affiliation{Department of Physics and Astronomy, Seoul National University, Seoul 08826, Korea}
\affiliation{Center for Theoretical Physics, Seoul National University, Seoul 08826, Korea}
\author{Cheol-Hwan Park}
\email{cheolhwan@snu.ac.kr}
\affiliation{Center for Correlated Electron Systems, Institute for Basic Science, Seoul 08826, Korea}
\affiliation{Department of Physics and Astronomy, Seoul National University, Seoul 08826, Korea}
\affiliation{Center for Theoretical Physics, Seoul National University, Seoul 08826, Korea}




\date{\today}

\begin{abstract}
Here, a magnetic model with an unprecedentedly large number of parameters was determined from first-principles calculations for transition-metal phosphorus trisulfides (TMPS$_3$'s), which reproduced the measured magnetic ground states of bulk TMPS$_3$'s.
Our Monte Carlo simulations for the critical temperature, magnetic susceptibility, and specific heat of bulk and few-layer TMPS$_3$'s agree well with available experimental data and show that the antiferromagnetic order of FePS$_3$ and NiPS$_3$ persists down to monolayers.
Remarkably, the orbital polarization, which was neglected in recent first-principles studies, dramatically enhances the magnetic anisotropy of FePS$_3$ by almost two orders of magnitude.
A recent Raman study [K. Kim et al., Nat. Commun. {\bf 10}, 345 (2019)] claimed that magnetic ordering is absent in monolayer NiPS$_3$ but simultaneously reported a strong two-magnon continuum; we show that the criterion used to judge magnetic ordering there is invalid in monolayer NiPS$_3$, thus providing an understanding of the two seemingly contradictory experimental results.
The rich predictions on the magnetic susceptibility and specific heat of few-layer FePS$_3$ and NiPS$_3$ await immediate experimental verifications.
\end{abstract}

\maketitle


Since the discovery of ferromagnetic materials Cr$_2$Ge$_2$Te$_6$~\cite{gongN2017} and CrI$_3$~\cite{huangN2017}, there have been a plethora of studies on two-dimensional (2D) ferromagnetic materials.
Recently, the research focus of 2D magnetic materials are moving from ferromagnetic 2D materials toward antiferromagnetic 2D materials.
As Cr$_2$Ge$_2$Te$_6$ and CrI$_3$ have played the role of first and representative 2D ferromagnetic materials, few-layer transition-metal phosphorus trisulfides (TMPS$_3$’s) were synthesized first~\cite{leeAM2016,duAN2016,kuoSR2016} and are playing the similar role for 2D antiferromagnetic materials.
A series of Raman studies on the stability of magnetic orders in atomically thin TMPS$_3$ compounds was published recently~\cite{leeNL2016,kimNC2019,kim2M2019,limCAP2021}, which have so far been very highly cited reflecting that these first, representative 2D antiferromagnetic materials are of great interest.
More recently, the strong coupling between the magnetic order and lights in FePS$_3$ and NiPS$_3$ is drawing huge attention~\cite{kangN2020,hwangboNN2021,zhangNL2021}.

Unlike the case of 2D ferromagnetic materials, whose magnetic properties and anisotropies are relatively well-known, there have been no satisfactory studies on these subjects for 2D antiferromagnetic materials so far.
So, it is very timely and of broad importance to understand the magnetic properties, especially magnetic anisotropies, of the first and representative 2D antiferromagnetic materials from theory and first-principles calculations. 

Moreover, there is more interesting physics in these compounds than 2D ferromagnetic materials from the viewpoint of the ``magnetism in 2D materials.''
According to the Mermin-Wagner theorem~\cite{merminPRL1966}, the magnetic ordering in 2D materials depends critically on the kind of detailed interactions among the magnetic ions: the magnetic ordering at a finite temperature is absent but there is the so-called Berezinskii-Kosterlitz-Thouless transition~\cite{berezinskiiSJETP1972,kosterlitzJPCSSP1973} in 2D XXZ systems.
As to investigating this peculiar character of the 2D XXZ model, few-layer NiPS$_3$'s were considered an ideal testbed because their antiferromagnetic ordering is believed to be described by the XXZ model~\cite{joyPRB1992}.
In a recent Raman study~\cite{kimNC2019}, magnetic ordering at a finite temperature of monolayer NiPS$_3$ was reported to be absent;
this conclusion, which was based on their observation of complete absence of the frequency shift of certain peaks in the Raman spectra of monolayer NiPS$_3$, seemed to agree with the aforementioned belief: ``NiPS$_3$ is (or at least is very close to) an ideal XXZ system so its magnetic ordering is critically affected by its thickness.''
In this perspective, however, it still remained a puzzle that prominent two-magnon signals were also observed in the same Raman spectra of monolayer NiPS$_3$ at finite temperatures.  
If the emergence of strong two-magnon continuum in monolayer NiPS$_3$ is a clear evidence of magnetic ordering at a finite temperature, then why were the frequency shifts of certain Raman peaks, which accompany a magnetic phase transition in the case of bulk NiPS$_3$, not detected in the case of monolayer NiPS$_3$ at all although they were all detected in the cases of bilayer or thicker samples as reported in ref~\citenum{kimNC2019}? Considering that ref~\citenum{kimNC2019} has significantly affected the following studies from the viewpoint of applying the Mermin-Wagner theorem to real 2D materials, understanding these seemingly contradictory experimental results is clearly of fundamental importance.

In this paper,
we demonstrate that a magnetic model that accurately describes the basic magnetic properties of all three \tmps{}'s (MnPS$_3$, FePS$_3$, and NiPS$_3$)---ordering pattern, magnetic anisotropy, $T_\mathrm{N}$, magnetic susceptibility, and specific heat---can be constructed from density functional theory+$U$ (DFT+$U$) calculations using fully-relativistic pseudopotentials.
The magnetic model presented here includes several factors such as the magnetic dipolar anisotropy, orbital degrees of freedom, and interlayer exchange interactions as well as anisotropies in (both intralayer and interlayer) exchange interactions, and single-ion anisotropy.
Based on Monte Carlo simulations, we show those factors sometimes individually and sometimes as a whole play a crucial role in explaining the magnetic properties of \tmps{}'s and their thickness dependence.

In the case of TMPS$_3$, the magnetic models with up to 10 parameters were considered~\cite{vaclavkova2M2020,olsenJPAP2021}, but still there is no comprehensive magnetic model for TMPS$_3$ that deals with various kinds of magnetic anisotropies and interlayer exchange interactions simultaneously.
Even in a very recent examination of the magnetic anisotropy of CrI$_3$---arguably the most famous 2D ferromagnetic material that has been thoroughly studied over the past few years---less than 10 (symmetrically non-equivalent) parameters were considered~\cite{bacaksizPRB2021}.  
Not to mention that a careful consideration of the tensorial form of all the magnetic interactions restricted by the symmetry ($C2/m$) of TMPS$_3$ was taken here, just the number, 73 parameters for each TMPS$_3$, is unprecedentedly large compared to the usual practice of studying magnetic interactions.
Here, we demonstrate that our comprehensive magnetic models for TMPS$_3$, which took an unprecedented amount of work with respect to the standard of this field, were indeed necessary for us to reach an agreement with available experimental results and several new predictions that would attract the attention of the experimentalists.

Remarkably, we show that the orbital polarization, which was neglected in recent first-principles studies~\cite{nauman2M2021,olsenJPAP2021}, enhances the magnetic anisotropy of FePS$_3$ by almost two orders of magnitude and hence is of crucial importance in explaining the experimental results on its magnetic properties.
Moreover, our calculation results predicted magnetic ordering at a finite temperature for monolayer NiPS$_3$, whose magnetic ordering was reported to be absent in a recent Raman study~\cite{kimNC2019}.
Importantly, we show that the criterion used in the Raman study to detect magnetic ordering does not work in the case of monolayer NiPS$_3$ and prove the conclusion of the Raman study that there is no magnetic ordering in monolayer NiPS$_3$~\cite{kimNC2019} to be wrong.

\begin{figure}
\begin{center}
\includegraphics[width=\columnwidth]{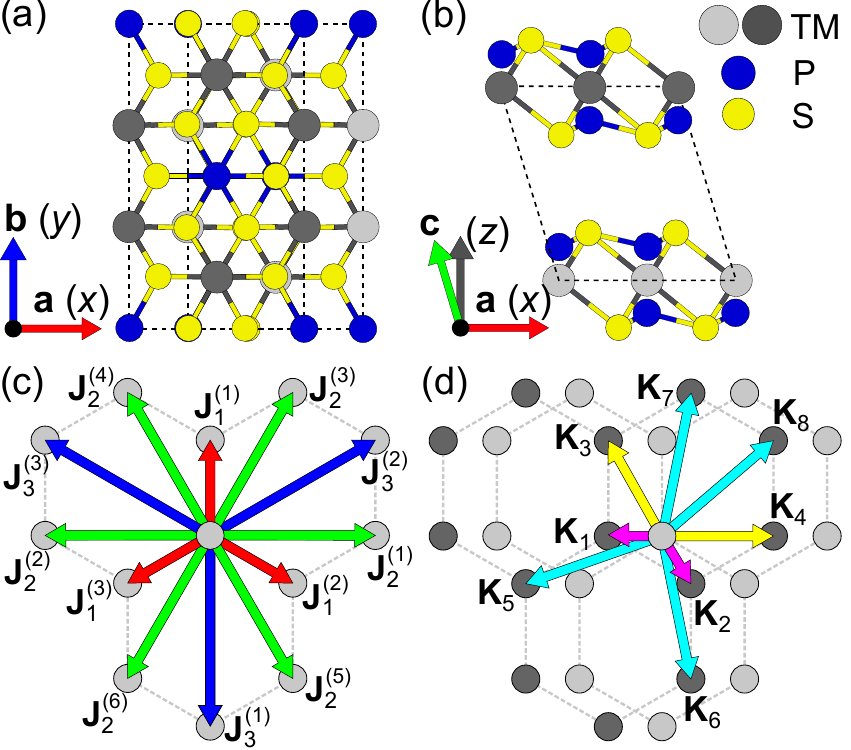}
\end{center}
\caption{
(a) Top view and (b) side view of the crystal structure of bulk TMPS$_3$'s. 
The global Cartesian axes
(\textit{x}, \textit{y} and \textit{z})
are defined such that the \textit{x} and \textit{y} directions coincide with 
the crystallographic direction \textbf{a} and \textbf{b}, respectively.
(c) Intralayer and
(d) interlayer exchange interactions considered.
}
\label{fig:structure_exchange}
\end{figure}

TMPS$_3$'s are layered materials (Figures~\ref{fig:structure_exchange}a and b) with antiferromagnetic orders that vary with their transition metal (TM = Mn, Fe, and Ni) element~\cite{brecIilm1986}.
The first constituent of our magnetic model for TMPS$_3$'s is the intralayer and interlayer exchange interactions which are written as
\begin{equation}
    H^{\mathrm{intra}}_{\mathrm{exch}} = 
    \frac{1}{2} \sum_{i=1}^{N} \sum_{n=1}^3 \sum_{a=1}^{M_n}
    {\mat{S}_i^{\mathsf{T}} \,  \mat{J}_{n}^{(a)} \, \mat{S}_{j(i,\,\mat{J}_n^{(a)})}} \\
\label{eqn:H_intra}
\end{equation}
and
\begin{equation}
    H^{\mathrm{inter}}_{\mathrm{exch}} =
    \frac{1}{2} \sum_{i=1}^{N} \sum_{n=1}^8
    {\mat{S}_{i}^{\mathsf{T}}\,\mat{K}_{n}\,\mat{S}_{j(i,\,\mat{K}_n)}},
\label{eqn:H_inter}
\end{equation}
respectively,
where the site index $i$ runs over $N$ magnetic atoms in the system,
$\mat{S}_i$ is the spin at the $i$-th site,
$\mat{J}_n^{(a)}$ and $\mat{K}_n$ are symmetric matrices representing anisotropic intralayer and interlayer exchange interactions, respectively, 
and $\mat{S}_{j(i,\,\mat{J}_n^{(a)})}$ and $\mat{S}_{j(i,\,\mat{K}_n)}$ denote 
the spins linked with $\mat{S}_i$
by $\mat{J}_n^{(a)}$ and $\mat{K}_n$, respectively 
(Figures~\ref{fig:structure_exchange}c,d).
$\mat{J}_n^{(a)}$'s corresponding to
the nearest neighbors $(\mat{J}_1^{(1\text{--}3)})$, second-nearest neighbors ($\mat{J}_2^{(1\text{--}6)}$), and third-nearest neighbors ($\mat{J}_3^{(1\text{--}3)}$) were included.
In the case of $\mat{K}_n$'s,
eight different interlayer exchange paths $(\mat{K}_{1\text{--}8})$ were considered.

The next is the single-ion anisotropy ($H_\mathrm{SIA}$):
\begin{equation}
    H_{\mathrm{SIA}} =
    \sum_{i=1}^{N} {\mat{S}_{i}^{\mathsf{T}}\,\mat{D}\,\mat{S}_{i}},
\label{eqn:H_SIA}
\end{equation}
where $\mat{D}$ is a symmetric matrix.
In previous studies, a simpler version---diagonal $\mat{D}$ with $D_{xx}$ = $D_{yy}$ $\ne$ $D_{zz}$---has been used~\cite{joyPRB1992,lanconPRB2018}.

In \feps{}, 
a coupling between orbital and spin polarizations ($LS$ coupling) leads to a strong magnetic anisotropy whose magnitude is about a few tens of meV~\cite{joyPRB1992}, estimated from experimental results using crystal field theory.
We therefore considered the following on-site contribution:
\begin{equation}\label{eqn:H_LS}
H_{LS} =  
\lambda{} \sum_{i=1}^{N} L_{iz} S_{iz},
\end{equation}
where $\lambda$ represents the $LS$ coupling strength, and $L_{iz}$, which can be either $1$ or $-1$, is the magnetic quantum number for the orbital polarization of an Fe ion.

All the parameters of $H_\text{exch}^\text{intra}$, $H_\text{exch}^\text{inter}$, $H_\text{SIA}$, and $H_{LS}$ (eqs~\ref{eqn:H_intra}--\ref{eqn:H_LS}) were obtained based on DFT+$U$ calculations;
see Supporting Information (SI) for calculation details of the anisotropic magnetic models.

Because the usual DFT(+$U$) methods cannot properly account for the dipolar coupling between local magnetic moments~\cite{pellegriniPRB2020},
we included the dipolar anisotropy ($H_{\rm dip}$) as the last constituent of our magnetic model:
\begin{equation}
\label{eqn:H_dip}
H_{\mathrm{dip}}
=
\frac{1}{2} \sum_{i=1}^N \sum_{j=1, j\ne{}i}^N {}
\left[
\frac{\mat{m}_i \cdot \mat{m}_j}{r_{ij}^3} - 
\frac{3(\mat{m}_i \cdot \mat{r}_{ij})(\mat{m}_j \cdot \mat{r}_{ij})}{r_{ij}^5}
\right],
\end{equation}
where $\mat{m}_i$ and $\mat{m}_j$ are the local magnetic moments at the $i$-th and $j$-th sites, 
and $r_{ij}$ is the distance between $\mat{m}_i$ and $\mat{m}_j$;
see SI for computational details of $H_{\mathrm{dip}}$.

\begin{figure}
\begin{center}
\includegraphics[width=\columnwidth]{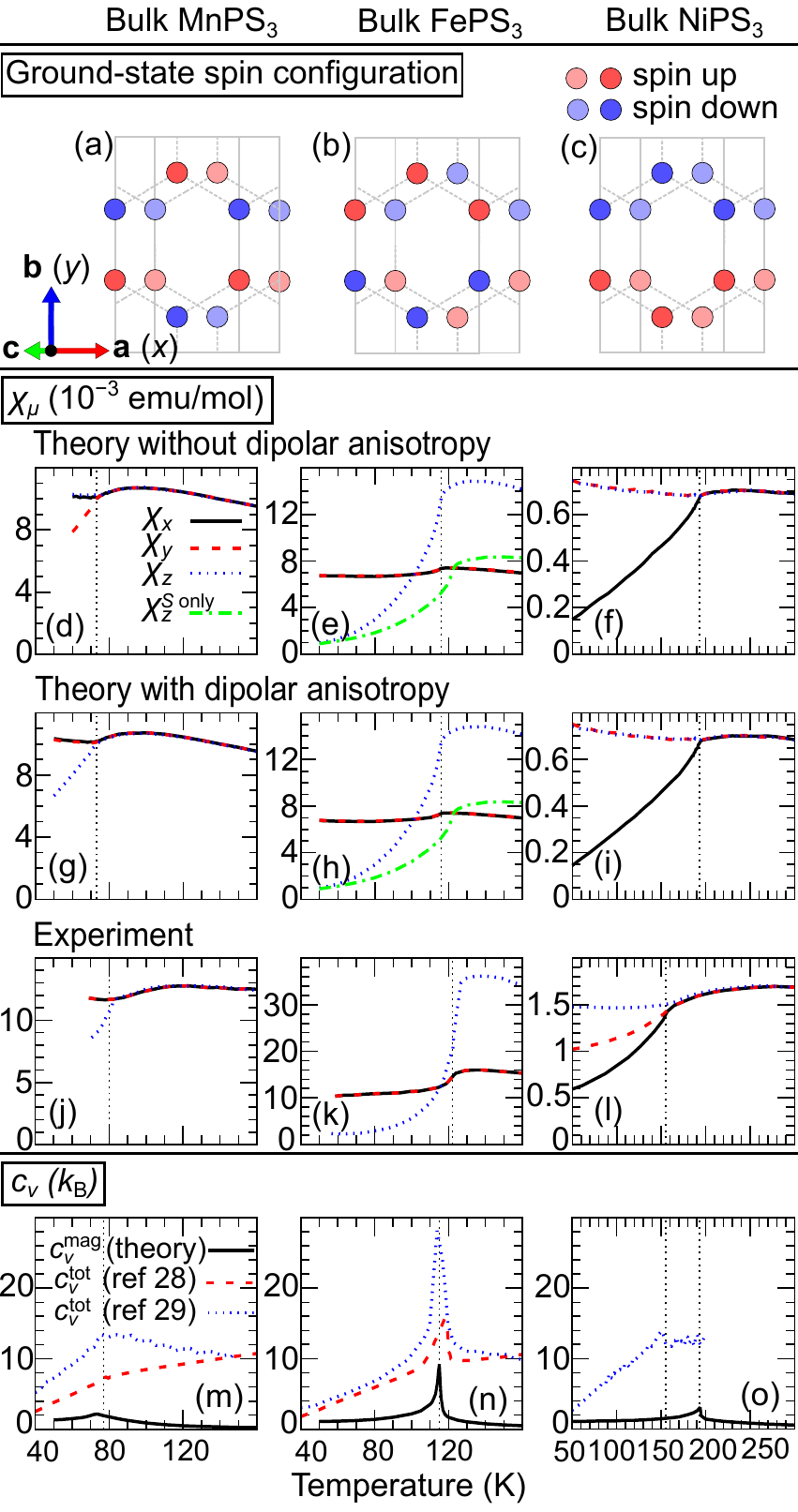}
\end{center}
\caption{
(a)--(c) The calculated ground-state spin-ordering pattern of bulk \tmps{}'s.
(d)--(l) The magnetic susceptibility of bulk TMPS$_3$'s.
The upper [(d)--(f)] and middle [(g)--(i)] panels show our 
calculation results, respectively, without and with considering 
the dipolar anisotropy.
The lower [(j)--(l)] panels show the measured magnetic susceptibilities for bulk \mnps{} 
and \feps{}~\cite{joyPRB1992} and for bulk \nips{}~\cite{wildesPRB2015}.
In the case of \feps{} [(e) and (h)], $\chi_z$
includes the effects of orbital polarizations
(see also SI for computational details).
(m)--(o) The calculated (magnetic) specific heat $c_v^\mathrm{mag}$ of bulk TMPS$_3$'s and
the experimental data for the (total) specific heat $c_v^\textrm{tot}$~\cite{takanoJMMM2004,siskinsACP2019}.
The vertical dashed lines show the estimated critical temperatures.
}
\label{fig:chi_bulk_comp}
\end{figure}

Remarkably, the ground-state ordering patterns and the directions of spin moments of all three compounds obtained from our first-principles calculations (Figures~\ref{fig:chi_bulk_comp}a--c) agree perfectly with those obtained from neutron scattering experiments~\cite{kurosawaJPSJ1983,lanconPRB2016,wildesPRB2015,lanePRB2020}.

Table~\ref{tab:mag_aniso} summarizes the contributions of $H_{\mathrm{exch}}^{\mathrm{intra}}$, $H_{\mathrm{exch}}^{\mathrm{inter}}$, $H_{\mathrm{SIA}}$, $H_{LS}$ and $H_{\mathrm{dip}}$ to the magnetic anisotropy of bulk \tmps{}.
(i) In the case of \mnps{},  $H_{\mathrm{dip}}$ and $H_{\mathrm{SIA}}$ dominantly contribute to the magnetic anisotropy.
These two contributions are of opposite (easy-axis and easy-plane, respectively) types and their sum results in a small easy-axis anisotropy (12.7\,$\muup$eV) along $z$.
Similar conclusions were drawn from previous experiments~\cite{wildesJMMM2007, hicksJMMM2019}.
(ii) In the case of \feps{}, $H_{LS}$ dominantly gives rise to a strong easy-axis anisotropy (20.6\,meV)
along $z$.
The calculated $LS$ coupling strength $\lambda$ (10.2\,meV) matches well with the $\lambda$ obtained from measured paramagnetic susceptibility (11.1--11.5\,meV)~\cite{joyPRB1992}.
(iii) In the case of \nips{}, the contributions of $H_{\mathrm{SIA}}$, $H_{\mathrm{exch}}^{\mathrm{intra}}$ and $H_{\mathrm{dip}}$ are similar in magnitude.
Their sum results in an easy-axis anisotropy (112.5\,$\muup$eV) along $x$.
While some studies claimed that \nips{} is best described by an XXZ model with easy-plane anisotropy~\cite{joyPRB1992, kimNC2019}, our calculation results support the neutron scattering studies reporting
an easy-axis anisotropy of a few hundred $\muup$eV~\cite{lanconPRB2018} and the ordering direction of almost along $x$~\cite{wildesPRB2015}.

\begin{table}
\begin{center}
\begin{tabular}
{p{2.8em}p{1em}>{\raggedleft\arraybackslash}p{3em}*{4}{>{\raggedleft\arraybackslash}p{2.8em}}>{\raggedleft\arraybackslash}p{3em}p{1em}}
\hline\hline
& $\mat{n}$ & $H_{LS}$ & $H_{\rm SIA}$ & $H_{\rm exch}^{\rm intra}$ & $H_{\rm exch}^{\rm inter}$ & $H_{\rm dip}$ & $H_{\rm tot}$ & \\
\hline
\multirow{2.8}{*}{\mnps{}}
& $x$ & 0 & $-$13.5 & $-$1.4 & 0.1 & 18.6 & 3.8 & \\
& $y$ & 0 & $-$14.3 & $-$1.4 & $-$0.1 & 17.6 & 1.6 & \\
& $z$ & 0.2 & 27.7 & 3.0 & 0.0 & $-$36.2 & $\mathbf{-5.3}$ & \textbf{(\textit{z})} \\
\hline
\multirow{2.8}{*}{\feps{}}
& $x$ & 0 & $-$10 & 404 & 15.6 & $-$44.4 & 365 & \\
& $y$ & 0 & $-$124 & $-$120 & $-$1.0 & 33.8 & $-$211 & \\
& $z$ & $-$20416 & 134 & $-$284 & $-$14.5 & 10.6 & $\mathbf{-20567}$ & \textbf{(\textit{z})} \\
\hline
\multirow{2.8}{*}{\nips{}}
& $x$ & 0 & $-$38.7 & $-$21.1 & $-$2.9 & $-$11.9 & $-\mathbf{74.6}$ & \textbf{(\textit{x})} \\
& $y$ & 0 & $-$21.0 & 45.7 & $-$0.2 & 12.4 & 36.9 & \\
& $z$ & $-$15 & 59.8 & $-$24.6 & 3.2 & 0.6 & 39.0 & \\
\hline\hline
\end{tabular}
\caption{
Magnetic anisotropy of bulk TMPS$_3$'s.
The total energy ($H_\text{tot}$, the sum of eqs~\ref{eqn:H_intra}--\ref{eqn:H_dip}) per transition metal ion at a given ordering direction $\mat{n}$ ($x$, $y$ and $z$ in Figures~\ref{fig:structure_exchange}a,b) is shown in units of $\muup$eV.
The parenthesis in the last column shows the easy axis.
The contributions of intralayer ($H_\text{exch}^\text{intra}$) and interlayer ($H_\text{exch}^\text{inter}$) exchange interactions, single-ion anisotropy ($H_\text{SIA}$), $LS$ coupling ($H_{LS}$), and dipolar anisotropy ($H_\text{dip}$) (eqs~\ref{eqn:H_intra}--\ref{eqn:H_dip}) are also shown.
}
\label{tab:mag_aniso}
\end{center}
\end{table}

Table~\ref{tab:intra_exch} summarizes the isotropic intralayer exchange interaction $J_n^{(a)} (= 1/3\,{\rm Tr}\,\mat{J}_n^{(a)}$) obtained from our calculations (here, the subscript $n$ in $J_n^{(a)}$ denotes the $n$-th nearest neighbor). 
For each $n$, there are two groups of $J_n^{(a)}$, e.g. $J_1^{(1)}$ and $J_1^{(2,3)}$, which are distinguished by the $C2/m$ symmetry of the crystal.
These two groups should have different values for the isotropic exchange interactions in principle.
However, since the distortion from a perfect honeycomb lattice is small, the difference between the two different values of $J_n^{(a)}$ turns out to be small for all $n$'s.

\begin{table}
\begin{center}
\begin{tabular}{p{3em}*{6}{>{\raggedleft\arraybackslash}p{3em}}}
\hline\hline
& $J_1^{(1)}$ & $J_1^{(2,3)}$ & $J_2^{(1,2)}$ & $J_2^{(3\text{--}6)}$ & $J_3^{(1)}$ & $J_3^{(2,3)}$ \\
\hline
\mnps{} & 1.06 & 1.18 & 0.07 & 0.07 & 0.72 & 0.69 \\
\feps{} & $-$1.36 & $-$1.36  & 0.11 & 0.12 & 2.37 & 2.40 \\
\nips{} & $-$5.21 & $-$5.34 & $-$0.23 & $-$0.20 & 27.4 & 28.2 \\
\hline\hline
\end{tabular}
\caption{
Isotropic intralayer exchange interaction $J_n^{(a)} = 1/3\,\mathrm{Tr}\,\mathbf{J}_n^{(a)}$ in units of meV (see Figure~\ref{fig:structure_exchange}c also).
}
\label{tab:intra_exch}
\end{center}
\end{table}

In the case of \mnps{} and \feps{}, the calculated $J_n^{(a)}$'s are in good agreement with neutron scattering experiments~\cite{lanconPRB2018}. 
In the case of \nips{}, $J_3^{(1\text{--}3)}$ obtained from our calculations are larger than the value fitted with a model to experimental results, $J_3$ = 13.8\,meV~\cite{lanconPRB2018}. (Here, we doubled this value in order to compensate the difference in the definition of $J_3$ in ref~\citenum{lanconPRB2018} and in this study.)
A rough estimate of the spin-wave gap in \nips{} is $\sim$ $2S\sqrt{3 \Delta J_3}$~\cite{lanconPRB2018}, where $\Delta$ is the single parameter that represents the total magnetic anisotropy energy per transition metal ion.
Putting $J_3$ = 28\,meV (Table~\ref{tab:intra_exch}), and $\Delta$ = 0.11\,meV (the magnitude of the easy-axis anisotropy in \nips{}, see Table~\ref{tab:mag_aniso}) into the expression for the spin-wave gap yields 6.1\,meV, which is very close to the measured spin-wave gap (7\,meV)~\cite{lanconPRB2018}.
$J_3$ used in the experimental study is an intermediate parameter, which is indirectly estimated by a fitting procedure. 
On the other hand, the spin-wave gap is a directly measured quantity from neutron scattering spectra, and our calculations explain this measured gap quite well.

The classical Monte Carlo method, describing the magnetic ordering of ferromagnetic 2D materials well~\cite{torelli2M2019, olsenMC2019}, was employed in our study on the temperature and thickness dependence of the thermodynamic quantities of \tmps{}'s;
see SI for discussions on the validity of using the classical Monte Carlo simulation.
Figures~\ref{fig:chi_bulk_comp}d--l show the magnetic susceptibility of bulk \tmps{}'s.
The calculated $T_\text{N}$'s of \mnps{} and \feps{} are 72\,K and 116\,K, respectively, in good agreement with experiments (80\,K and 120\,K, respectively)~\cite{brecIilm1986, joyPRB1992}.
In the case of \mnps{} (Figure~\ref{fig:chi_bulk_comp}g), not only the $T_\text{N}$ but also the anisotropy of the magnetic susceptibility (the differences between $\chi_{x}$, $\chi_{y}$ and $\chi_{z}$) agrees with the experimental data 
(Figure~\ref{fig:chi_bulk_comp}j).
As can be seen in Figures~\ref{fig:chi_bulk_comp}d and g, $H_\text{dip}$ plays a crucial role in reproducing the anisotropy of the magnetic susceptibility below $T_\text{N}$.
Quite remarkably, in the case of \feps{} (Figure~\ref{fig:chi_bulk_comp}h), the calculated magnetic susceptibilities reproduce the significant difference between $\chi_z$ and $\chi_{x,y}$ over all temperatures, in agreement with the experiment (Figure~\ref{fig:chi_bulk_comp}k).
Here, we emphasize the importance of including the orbital degrees of freedom ($L_{iz}$ in eq~\ref{eqn:H_LS}) in calculating the magnetic susceptibility:
the significant enhancement of $\chi_z$ shown in Figures~\ref{fig:chi_bulk_comp}e and h (see also SI for how the orbital contribution was considered in our Monte Carlo simulations).

In the case of \nips{} (Figure~\ref{fig:chi_bulk_comp}i), although the calculated $T_\text{N}$ (193\,K) is 38\,K higher than the experimental value (155\,K)~\cite{joyPRB1992,wildesPRB2015},
it is in much better agreement with experiments than previous theoretical estimations (353\,K and 560\,K from refs~\citenum{lanconPRB2018} and \citenum{chittariPRB2016}, respectively).
More importantly, our calculations reproduce several key features observed in experiments:
(i) the susceptibility is very isotropic ($\chi_x = \chi_y = \chi_z$) if $T > T_\text{N}$, (ii) there is a broad peak far above $T_\text{N}$ at $T\sim$250\,K, and (iii) $\chi_x$ is the smallest if $T < T_\text{N}$,
indicating that the antiferromagnetic order is almost along $x$~\cite{wildesPRB2015, lanconPRB2018}.

Figures~\ref{fig:chi_bulk_comp}m--o show the magnetic specific heat ($c_v^\mathrm{mag}$) of bulk \tmps{}.
The peak in the $c_v^\text{mag}$ of \feps{} is much sharper than those in the other cases, in good agreement with experiments~\cite{takanoJMMM2004, siskinsACP2019},
which is easily understood in that the magnetic anisotropy of \feps{} is orders of magnitude larger than that of the other \tmps{}'s (Table~\ref{tab:mag_aniso}).
We note that the measured total specific heat ($c_v^\mathrm{tot}$) for bulk \tmps{} shown in Figures~\ref{fig:chi_bulk_comp}m--o includes the contributions arising from phonons and other possible noises
and that the discrepancies between the different experimental results call for further experimental studies guided by our paper.

\begin{figure}
\begin{center}
\includegraphics[width=\columnwidth]{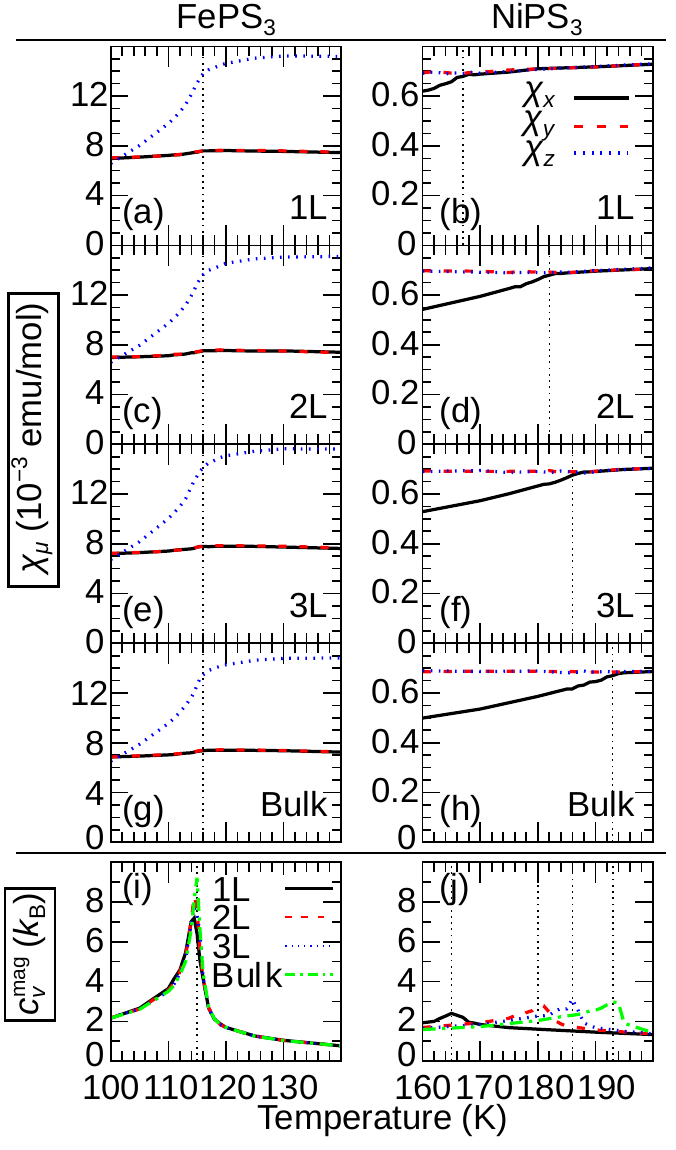}
\end{center}
\caption{
(a)--(h) The magnetic susceptibility of bulk and few-layer TMPS$_3$'s. 
(i)--(j) The specific heat arising from magnetic moments $c_v^\mathrm{mag}$ of bulk and few-layer TMPS$_3$'s.
The vertical dashed lines show the estimated N\'eel temperatures.
}
\label{fig:chi_thickness}
\end{figure}

Now we discuss how the $T_\text{N}$ and magnetic susceptibility change with the number of layers in few-layer FePS$_3$ and NiPS$_3$.
In Figure~\ref{fig:chi_thickness},
we first notice that the magnetic phase transition exists in the monolayer (1L) limit.
The thickness dependence of the $T_\text{N}$ significantly varies with the transition metal element.
In the case of \feps{}, the $T_\text{N}$ is (almost) independent of the number of layers, also in good agreement with experiments~\cite{leeNL2016}.
The interlayer exchange interactions in 
\feps{} (see SI Table~S1) are orders of magnitude smaller than the intralayer ones (Table~\ref{tab:intra_exch}) and the $LS$ coupling (see Table~\ref{tab:mag_aniso});
this explains why the $T_\text{N}$ hardly changes with the thickness.

By contrast, $T_\text{N}^\mathrm{bulk} - T_\text{N}^\mathrm{1L}$ of \nips{} (28\,K) is much larger than that of 
\feps{} (0\,K);
this is easily explained by the unusually large interlayer exchange interactions of \nips{},
especially the second-nearest-neighbor ones ($K_{3,4}$ =$1/3\,\mathrm{Tr}\,\mat{K}_{3,4}$ $\sim$ 1.64\,meV, see SI Table~S1), which are as large as 30\,\% of the nearest-neighbor intralayer exchange interactions ($J_1^{(1\text{--}3)}$ $\sim$ $-$5.3\,meV, see Table~\ref{tab:intra_exch}).

Notably, we predict a finite (and also considerably large) $T_\mathrm{N}$ for monolayer \nips{} (Figures~\ref{fig:chi_thickness}b and j),
which indicates that the magnetic order persists in the genuine 2D limit of \nips{}.
On the other hand, 
a recent Raman study drew exactly the opposite conclusion that the magnetic order is absent in monolayer \nips{} (but is present in bilayer or thicker samples)~\cite{kimNC2019},
which was largely based on their observation that the frequencies of certain Raman peaks (denoted as P$_2$ in ref~\citenum{kimNC2019}, the $E_g$ modes at $\sim$180\,cm$^{-1}$) did not budge with decreasing the temperature (far below the bulk $T_\mathrm{N}$) in monolayer samples
while noticeably shifted in thicker samples.

At the same time, ref~\citenum{kimNC2019} also reported the emergence of a strong two-magnon continuum in the Raman spectrum of monolayer \nips{}, which is a clear evidence of magnetic ordering, at similar temperatures as in thicker samples. 
Thus, the two seemingly contradictory experimental results have remained a puzzle so far. 

Remarkably, we found that
such a complete absence of the frequency shifts of P$_2$ peaks
in the Raman spectrum of {\it magnetically-ordered} monolayer \nips{} can be fully understood
if the effects of the three-fold-degenerate ground-state manifold of monolayer \nips{}, as in other honeycomb-lattice compounds~\cite{searsPRB2017}, is considered.
Our resolution of this important puzzle, together with the results of our calculations, fully explains the seemingly contradictory experimental results in ref~\citenum{kimNC2019} but, at the same time, disproves the conclusion therein.

In the previous Raman study~\cite{kimNC2019}, 
a pair of degenerate Raman peaks (labeled as P$_2$ in ref~\citenum{kimNC2019}), 
one of which was only visible in the parallel-polarization setup [$-z(xx)z$] and the other only in the cross-polarization setup [$-z(xy)z$], 
was considered an order parameter of bulk \nips{} because a frequency difference between the two Raman peaks (denoted as $\Delta$P$_2$ from now on) started to develop just below $T_\mathrm{N}$ and increased further as the temperature decreased while $\Delta$P$_2$ was negligible above $T_\mathrm{N}$.  
The same phenomenon was observed from the bulk down to the bilayer.  
In monolayers on the other hand, $\Delta$P$_2$ remained zero at all temperatures.  
The result was then interpreted that the magnetic order is absent only in monolayer \nips{}.

However, we realized that $\Delta$P$_2$ cannot be measured from Raman experiments with the two (parallel- and cross-) polarization setups 
even if the magnetic order persists in monolayer \nips{}.  
A monolayer \nips{} has a three-fold rotation symmetry (the $D_{3d}$ point group) due to the lack of monoclinic stacking; 
consequently, its zigzag antiferromagnetic ground states are triply degenerated (imagine three different zigzag directions that can fit in a honeycomb lattice).  
An important note here is that all three degenerate states appear in a sample as magnetic domains with three different optical responses (this phenomenon is common in compounds with three-fold ground states interconnected by three-fold rotations~\cite{searsPRB2017}).  
The Raman responses of such domains should be averaged when we discuss the polarization-setup dependence of Raman peaks in monolayer \nips{}.

\begin{figure}
\begin{center}
\includegraphics[width=\columnwidth]{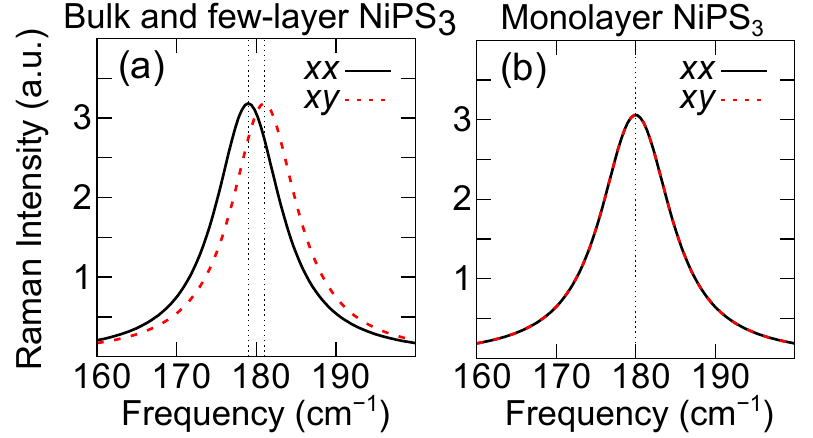}
\end{center}
\caption{
Simulated Raman spectra of \nips{} near the P$_2$ peaks for (a) few-layer and bulk \nips{} (the $C_{2h}$ point group) and (b) monolayer \nips{} (the $D_{3d}$ point group).  
It was assumed that the two peaks comprising P$_2$ are given by Lorentzians with the same height and width and their centers are separated by 2\,cm$^{-1}$.  
The half-width at half-maximum is 5\,cm$^{-1}$.  
These parameters were chosen to mimic the experimental results~\cite{kimNC2019}.  
In the legends, $xx$ and $xy$ represent the parallel- and cross-polarization setups, respectively.
}
\label{fig:intensity}
\end{figure}

The two P$_2$ peaks are assigned to the $E_g$ phonon mode in the $D_{3d}$ point group and their Raman tensors, 
which we will denote as $E_g^{(1)}$ and $E_g^{(2)}$, 
are respectively given by
\begin{equation}
\begin{pmatrix}
c & 0 & 0\\
0 & -c & d\\
0 & d & 0
\end{pmatrix}
\,\mathrm{and}\,
\begin{pmatrix}
0 & -c & -d\\
-c & 0 & 0\\
-d & 0 & 0
\end{pmatrix},
\label{eqn:tensor1}
\end{equation}
where $c$ and $d$ are real numbers.  
The Raman intensities in the parallel- and cross-polarization setups are proportional to the squares of the $xx$ and $xy$ components of the Raman tensors, respectively.  
The $E_g^{(1)}$ mode is therefore visible only in the parallel-polarization setup 
while the $E_g^{(2)}$ mode only in the cross-polarization setup.  
If magnetic ordering happens, a frequency difference between the two modes develops.  
Since the Raman peaks can be selectively captured (by adjusting the polarizations of incoming and outgoing lights),
$\Delta$P$_2$ can be measured accurately no matter how broad each of the two P$_2$ peaks is (see Figure~\ref{fig:intensity}a).

Note, however, that those Raman tensors are for one of the three zigzag orientations where the same-spin zigzag chains are aligned along the $x$ direction.  
The Raman tensors for the other zigzag orientations can be obtained by applying 120$^\circ$ and 240$^\circ$ rotations with respect to $z$ to the above expressions.  
In the case of the $E_g^{(1)}$ mode, the Raman tensors for all the different orientations are written as
\begin{equation}
\begin{pmatrix}
c & 0 & 0\\
0 & -c & d\\
0 & d & 0
\end{pmatrix}
\,,\,
\begin{pmatrix}
-\frac{c}{2} & -\frac{\sqrt{3}c}{2} & -\frac{\sqrt{3}d}{2}\\
-\frac{\sqrt{3}c}{2} & \frac{c}{2} & -\frac{d}{2}\\
-\frac{\sqrt{3}d}{2} & -\frac{d}{2} & 0
\end{pmatrix}
\,,\mathrm{and}\,
\begin{pmatrix}
-\frac{c}{2} & \frac{\sqrt{3}c}{2} & \frac{\sqrt{3}d}{2}\\
\frac{\sqrt{3}c}{2} & \frac{c}{2} & -\frac{d}{2}\\
\frac{\sqrt{3}d}{2} & -\frac{d}{2} & 0
\end{pmatrix}.
\end{equation}
Similarly, the three $E_g^{(2)}$ Raman tensors for the three different zigzag orientations read
\begin{equation}
\begin{pmatrix}
0 & -c & -d\\
-c & 0 & 0\\
-d & 0 & 0
\end{pmatrix}
\,,\,
\begin{pmatrix}
\frac{\sqrt{3}c}{2} & \frac{c}{2} & \frac{d}{2}\\
\frac{c}{2} & -\frac{\sqrt{3}c}{2} & \frac{\sqrt{3}d}{2}\\
\frac{d}{2} & \frac{\sqrt{3}d}{2} & 0
\end{pmatrix}
\,,\mathrm{and}\,
\begin{pmatrix}
-\frac{\sqrt{3}c}{2} & \frac{c}{2} & \frac{d}{2}\\
\frac{c}{2} & \frac{\sqrt{3}c}{2} & -\frac{\sqrt{3}d}{2}\\
\frac{d}{2} & -\frac{\sqrt{3}d}{2} & 0
\end{pmatrix}.
\end{equation}
Element-wise square-averaging of the Raman tensors over the three zigzag orientations now yields the following intensity matrices
\begin{equation}
\begin{pmatrix}
\frac{c^2}{2} & \frac{c^2}{2} & \frac{d^2}{2}\\
\frac{c^2}{2} & \frac{c^2}{2} & \frac{d^2}{2}\\
\frac{d^2}{2} & \frac{d^2}{2} & 0
\end{pmatrix}
\,\mathrm{and}\,
\begin{pmatrix}
\frac{c^2}{2} & \frac{c^2}{2} & \frac{d^2}{2}\\
\frac{c^2}{2} & \frac{c^2}{2} & \frac{d^2}{2}\\
\frac{d^2}{2} & \frac{d^2}{2} & 0
\end{pmatrix}
,
\end{equation}
respectively, for $E_g^{(1)}$ and $E_g^{(2)}$.  
Here, the $xx$ ($xy$) components of these two matrices give the domain-averaged Raman intensities for $E_g^{(1)}$ and $E_g^{(2)}$, respectively, in the parallel (cross) polarization setup.  
Remarkably, (i) these two matrices are exactly the same and (ii) their $xx$ and $xy$ components are also the same.  
An important point here is that there is certainly no polarization selection rule for the $E_g^{(1)}$ and $E_g^{(2)}$ modes if the averaging effects of the magnetic domains are taken into account: 
the two peaks comprising the P$_2$ peak are to be observed with equal intensities regardless of whether the parallel- or cross-polarization setup is used.

Such a disappearance of the polarization-setup dependence has a huge impact on detecting $\Delta$P$_2$ in practice; 
the impact is much more dramatic when the broadness of the two peaks comprising the P$_2$ peak is similar to or even larger than their frequency difference ($\Delta$P$_2$).  
If that condition is met, the two peaks {\it appear to be just one peak}.  
Unfortunately, the splitting in the peaks $\Delta$P$_2$ measured in ref~\citenum{kimNC2019} were at most $\sim$2\,cm$^{-1}$ (in few-layer and bulk \nips{}), 
and the half-width at half-maximum of the P$_2$ peaks in monolayer \nips{} were at least 5\,cm$^{-1}$.  
Given that condition, $\Delta$P$_2$ cannot be measured because of the large overlap between the peaks as shown in Figure~\ref{fig:intensity}b.

In summary, our anisotropic magnetic model with an unprecedentedly large number of parameters determined from extensive first-principles calculations and classical Monte Carlo simulations with special handling of the orbital polarization effects explained the magnetic structure, critical temperature, susceptibility, and specific heat of bulk and few-layer \tmps{}'s reported from experiments. Moreover, we have shown that the criterion used to judge magnetic ordering in the recent Raman study of NiPS$_3$~\cite{kimNC2019} is invalid in the case of monolayer samples and that magnetic ordering in NiPS$_3$ persists down to the monolayer limit.
Recently, direct measurements of the magnetic susceptibility of 2D materials were carried out~\cite{wongAM2019}.
Thus, our predictions on the thickness dependence of the thermodynamic quantities of few-layer \tmps{}'s can be verified immediately, which will allow a deeper understanding of magnetic ordering in atomically-thin 2D materials, a fundamental issue of the Mermin-Wagner theorem~\cite{merminPRL1966}.

\begin{acknowledgments}
We thank Je-Geun Park and Hyeonsik Cheong for many fruitful discussions during our previous collaborations and Choong H. Kim for helpful discussions. This work was supported by the Creative-Pioneering Research Program through Seoul National University, Korean NRF No-2020R1A2C1014760, and the Institute for Basic Science (No. IBSR009-D1). Computational resources were provided by KISTI Supercomputing Center (KSC-2020-INO-0078).
\end{acknowledgments}

\appendix
\setcounter{figure}{0}    
\setcounter{table}{0}    
\renewcommand{\thetable}{S\arabic{table}}
\renewcommand{\thefigure}{S\arabic{figure}}
\renewcommand{\theequation}{\thesection\arabic{equation}}

\section{Calculation details: anisotropic magnetic model from constrained DFT+$U$ calculations}
\subsection{Anisotropic magnetic model}
In this section, we describe how the anisotropic magnetic model $H_{\rm tot}$, which is given by
\begin{align}
H_{\rm tot} &= \frac{1}{2} \sum_{i=1}^{N} \sum_{n=1}^3 \sum_{a = 1}^{M_n} {\mat{S}_{i}^{\mathsf{T}}\,\mat{J}_{n}^{(a)}\,\mat{S}_{j(i,\mat{J}_n^{(a)})}} + \frac{1}{2} \sum_{i=1}^{N} \sum_{n=1}^8 {\mat{S}_{i}^{\mathsf{T}}\,\mat{K}_{n}\,\mat{S}_{j(i,\mat{K}_n)}} \nonumber \\
&+ \sum_{i=1}^{N} {\mat{S}_{i}^{\mathsf{T}}\,\mat{D}\,\mat{S}_{i}} + \sum_{i=1}^{N} \lambda L_{iz} S_{iz} + H_{\rm dip}\,, \label{supp_eqn:H_tot}
\end{align}
was determined from first-principles calculations.
The $C2/m$ symmetry of the crystal structure of \tmps{} was considered to restrict matrices $\mat{J}_n^{(a)}$, $\mat{K}_n$, and $\mat{D}$:
(i) $\mat{J}_1^{(1)}$, $\mat{J}_2^{(1,2)}$, $\mat{J}_3^{(1)}$, $\mat{K}_1$, $\mat{K}_4$, and $\mat{D}$ are of the form
\begin{equation}
\begin{pmatrix}
a & 0 & d\\
0 & b & 0\\
d & 0 & c
\end{pmatrix}\,,
\end{equation}
and (ii) the other matrices in eq~\ref{supp_eqn:H_tot} are of the form
\begin{equation}
\begin{pmatrix}
a & e & d\\
e & b & f\\
d & f & c
\end{pmatrix}\,,
\end{equation}
where $a$, $b$, $c$, $\cdots$, and $f$ are real numbers.

\subsection{Total energy mapping analysis}
To obtain the parameters in eq~\ref{supp_eqn:H_tot}, we first calculated the total energies of bilayer \tmps{}'s with different magnetic patterns and different collinear directions.
Figure~\ref{fig:energy_mapping_pattern} shows the collinear magnetic ordering patterns for which DFT+$U$ total energies were calculated.
We mention that the lowest-energy configurations obtained from our first-principles calculations coincide with the ordering patterns for all three bulk \tmps{} compounds reported from neutron scattering experiments: the patterns of Figures~\ref{fig:energy_mapping_pattern}c, g, and f for \mnps{}~\cite{kurosawaJPSJ1983}, \feps{}~\cite{kurosawaJPSJ1983,lanconPRB2016}, and \nips{}~\cite{wildesPRB2015}, respectively.
For each ordering pattern in Figure~\ref{fig:energy_mapping_pattern}, six different collinear directions were chosen, which means in total 132 total energy calculations were carried out for {\it each} \tmps{}.
With the $C2/m$ symmetry fully considered,
the total energies are then used to obtain 73 free parameters  for each \tmps{} using the least square method.
We note that for some of 22 ordering patterns depicted in Figure~\ref{fig:energy_mapping_pattern} the total energy was not employed as itself in the fitting procedure:
For the ordering patterns in Figures~\ref{fig:energy_mapping_pattern}e, j, and o,
the difference between the total energies of the two ordering patterns was used;
For the ordering patterns in Figure~\ref{fig:energy_mapping_pattern}p,
a specific combination of pluses and minuses of the total energies of the four ordering patterns was used as this particular choice gives the difference between $\mat{J}_3^{(1)}$ and $\mat{J}_3^{(2,3)}$. 
The spin directions of the transition metal ions were constrained to be along the collinear directions during the total energy calculations.
In the case of \mnps{} and \nips{}, the collinear direction $\mat{n}$ was chosen to be [1, 0, 0]$^\mathsf{T}$, [0, 1, 0]$^\mathsf{T}$, [0, 0, 1]$^\mathsf{T}$, [$\mathrm{sin}\,\theta$, 0, $\mathrm{cos}\,\theta$]$^\mathsf{T}$, [0, $\mathrm{sin}\,\theta$, $\mathrm{cos}\,\theta$]$^\mathsf{T}$, and [$\mathrm{cos}\,\phi$, $\mathrm{sin}\,\phi$, 0]$^\mathsf{T}$ (in Cartesian coordinates), where $\theta = 30^{\circ}$ and $\phi = 30^{\circ}$. 
In the case of \feps{}, a different choice for $\mat{n}$ was made to more accurately capture the magnetic anisotropy that strongly favors the $z$ direction: $\mat{n}$ = [$\mathrm{sin}\,\theta_1$, 0, $\mathrm{cos}\,\theta_1$]$^\mathsf{T}$, [0, $\mathrm{sin}\,\theta_1$, $\mathrm{cos}\,\theta_1$]$^\mathsf{T}$, [0, 0, 1]$^\mathsf{T}$, [$\mathrm{sin}\,\theta_2$, 0, $\mathrm{cos}\,\theta_2$]$^\mathsf{T}$, [0, $\mathrm{sin}\,\theta_2$, $\mathrm{cos}\,\theta_2$]$^\mathsf{T}$, and [$\mathrm{sin}\,\theta_2\,\mathrm{cos}\,\phi$, $\mathrm{sin}\,\theta_2\,\mathrm{sin}\,\phi$, $\mathrm{cos}\,\theta_2$]$^\mathsf{T}$, where $\theta_1 = 30^\circ$, $\theta_2 = 60^\circ$, and $\phi = 30^\circ$.

\subsection{Constrained DFT+$U$ calculations}
We carried out DFT+$U$ calculations using {\it our custom version} of Quantum ESPRESSO suite~\cite{giannozziJPCM2009}, in which we implemented a scheme that constrains the spin directions as proposed in ref~\citenum{maPRB2015} with our adaptations for the DFT+$U$ method.
In our implementation, 
the local magnetic moment at the $i$-th magnetic atom $\mathbf{m}_{i}$ was chosen to be
\begin{equation}
\mathbf{m}_{i} = \sum_{s_1 s_2} \sum_{m} \sum_{n\mathbf{k}} \left<\psi_{n\mathbf{k}}^{s_1} \big| \phi_{im}\right> {\boldsymbol\sigma}_{s_1s_2} \left<\phi_{im} \big| \psi_{n\mathbf{k}}^{s_2}\right>,
\end{equation}
where $s_1$ and $s_2$ are the spinor indices,
$\left| \psi_{n\mathbf{k}}^{s} \right>$ is the spin $s$ component of the Kohn-Sham spinor wavefunction with the band index $n$ and momentum $\mathbf{k}$,
$\left| \phi_{im} \right>$ is the $3d$ atomic orbital of the $i$-th magnetic atom with the orbital magnetic quantum number $m$,
and ${\boldsymbol\sigma}_{s_1s_2}$ is the $(s_1, s_2)$ component of the Pauli matrices ${\boldsymbol\sigma} = (\sigma^x, \sigma^y, \sigma^z)$.
Then the spin direction of the magnetic atom is constrained along a predefined direction $\mathbf{e}_{i}$ by minimizing the DFT total energy plus the penalty energy given as~\cite{maPRB2015}
\begin{equation}
    E_p = \sum_{i} \lambda_i \left( \left| \mathbf{m}_i \right| - \mathbf{e}_{i} \cdot \mathbf{m}_i \right),
\end{equation}
where $\lambda_i$ determines how close the constrained spin direction of $\mathbf{m}_i$ will be to $\mathbf{e}_i$.
In our calculations $\lambda_i$ was typically set to 1~Ry, which yielded well-converged local magnetic moments, $\mathbf{m}_i$'s, close to their target directions, $\mathbf{e}_i$'s.

\subsection{Further computational details}
The exchange-correlation functional within the local density approximation (LDA) parametrized by Perdew and Zunger~\cite{perdewPRB1981} was used to calculate the exchange-correlation energy. 
We employed a rotationally invariant LDA+$U$ method suggested by Liechtenstein {\it et\,al\,.}~\cite{liechtensteinPRB1995}. 
The effective Coulomb interaction $U$ was set to 6.5\,eV and 4\,eV for \nips{} and the other \tmps{}'s, respectively.
In determining the $U$ values,
we first tested several different $U$ values in the range 3--7\,eV and chose the values that provide reasonable estimates for the $T_\mathrm{N}$ of the bulk \tmps{} compounds.
We then carried out linear response calculations with the hp.x program~\cite{timrovPRB2018} in the Quantum ESPRESSO suite to see whether those empirical $U$ values could be supported by the well-established framework for computing $U$ values from first principles.
Our linear response calculations yielded 2.9\,eV, 3.9\,eV, and 6.6\,eV for the transition metal elements in \mnps{}, \feps{}, and \nips{}, respectively, which are in good agreement with our empirically chosen $U$ values.
In general, $U$ tends to increase across the row in the periodic table (from left to right) as the 3$d$ electrons get more localized~\cite{aryasetiawanPRB2006}.
In terms of $U$, therefore, CoPS$_3$ ($Z$=27 for Co) is expected to be somewhere between FePS$_3$ ($Z$=26 for Fe) and NiPS$_3$ ($Z$=28 for Ni);
that would explain the gap between the $U$ of FePS$_3$ and that of NiPS$_3$.

Optimized, norm-conserving pseudopotentials~\cite{hamannPRB2013, schlipfCPC2015} that include the relativistic effects were used to simulate the interactions between the nuclei and electrons. 
The kinetic energy cutoff of the plane-wave basis was set to 80\,Ry.

The atomic coordinates of a bilayer \tmps{} were adopted from the atomic coordinates of the bulk structure reported from an experimental study~\cite{ouvrardMRB1985}.
Because of the over-binding behavior of the LDA, we did not carry out further structural relaxations.
This over-binding behavior of the LDA has always been a notorious problem because a physical property of a material is often sensitive to its geometry. 
And when the LDA fails to provide a reasonable result for the structure, as it really does in the case of \tmps{} as well as in the cases of many other materials, there is no way but to resort to the experimental structure since a structural relaxation with the LDA plus a van der Waals correction is not a viable option since all the available van der Waals correction methods are designed to work with the GGA. (In Section~\ref{sec:GGA_NCM} of this Supporting Information, we show why we chose to use the LDA and not the GGA.)
Fortunately, keeping the experimental structure is known to give good results for the magnetic properties~\cite{solovyevPRB2015, fangPRL2004, yareskoPRB2009, jacobssonPRB2013, foyevtsovaPRB2013, lambrechtPRB2003}, which are the main focus of our study, and we therefore used the experimental structure.
To avoid spurious interactions between  periodic images, a vacuum layer with 18\,\AA{} thickness was placed between the adjacent periodic images. 

\begin{figure*}
\begin{center}
\includegraphics[width=\textwidth]{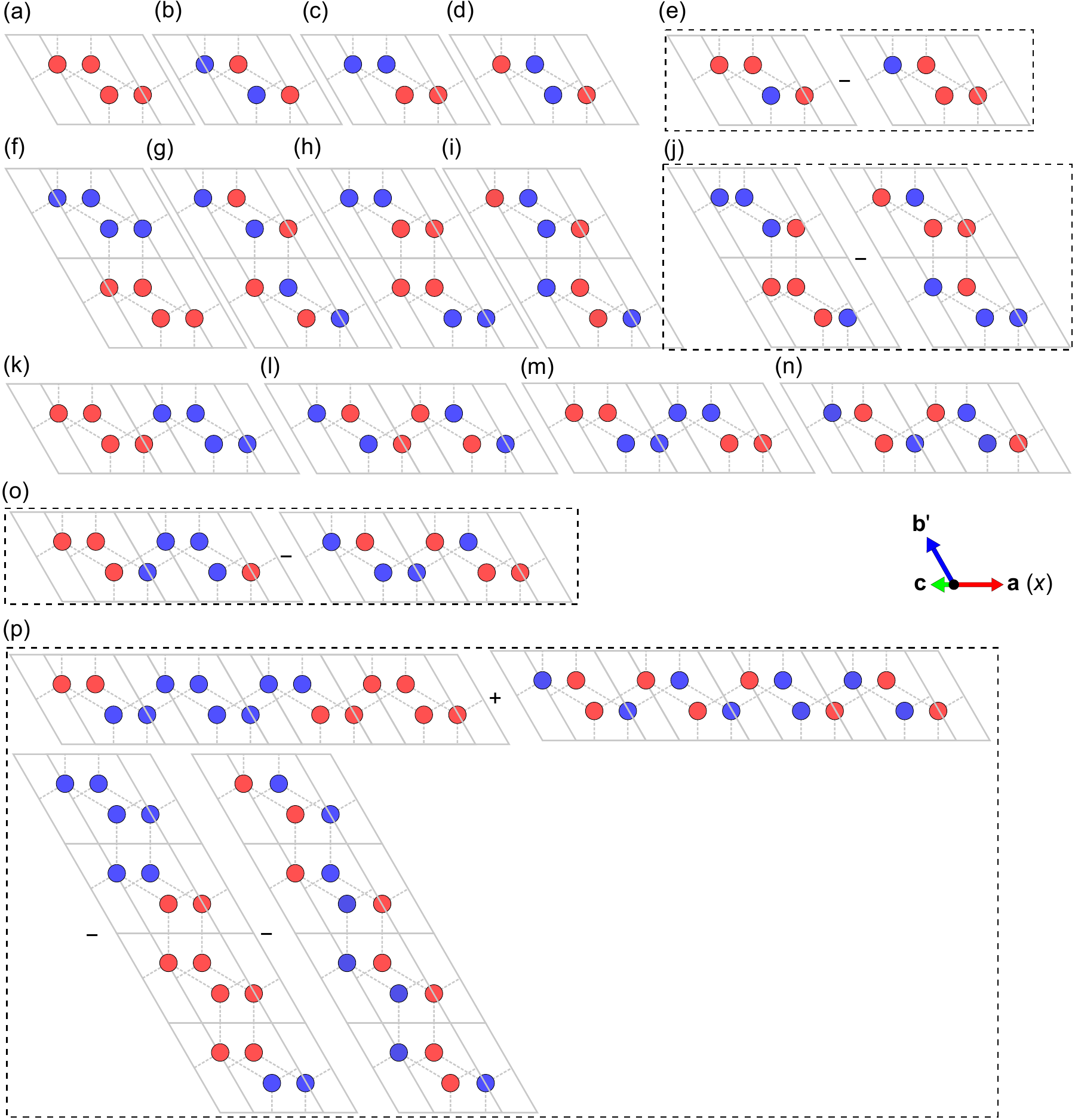}
\end{center}
\caption{
Top view of the collinear magnetic patterns of bilayer \tmps{} used to calculate the parameters of the anisotropic magnetic model (eq~\ref{supp_eqn:H_tot}). 
The red and blue disks represent the spin-up and spin-down transition metal ions, respectively, with respect to the collinear direction.
}
\label{fig:energy_mapping_pattern}
\end{figure*}

\begin{figure}
\begin{center}
\includegraphics[width=\columnwidth]{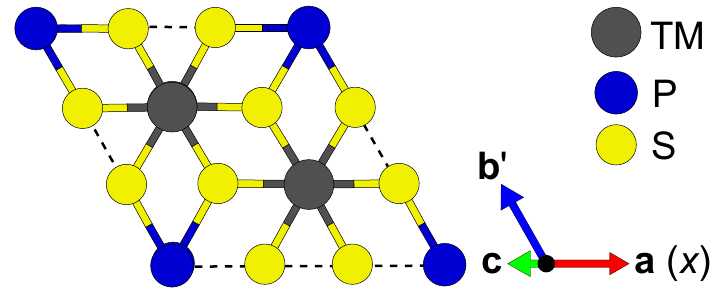}
\end{center}
\caption{
Top view of the primitive unit cell of nonmagnetic \tmps{}.
}
\label{fig:oblique_structure}
\end{figure}

We calculated the total energy for a supercell (Figure~\ref{fig:energy_mapping_pattern}) having $N_1 \times N_2 \times 1$ unit cells of nonmagnetic \tmps{} (Figure~\ref{fig:oblique_structure}), where ($N_1$, $N_2$) is (1, 1), (2, 1), (1, 2), (4, 1), and (1, 4).
To keep the $k$-point density in the reciprocal space for the calculations involving supercells of different size the same, we used an $8/N_1 \times 8/N_2 \times 1$ Monkhorst-Pack grid~\cite{monkhorstPRB1976} in the case of the $N_1 \times N_2 \times 1$-supercell calculations.

\subsection{Note on exchange-correlation functional with non-collinear magnetism}
\label{sec:GGA_NCM}
In this work, we did not use the generalized gradient approximation (GGA) because the GGA with noncollinear magnetism (GGA+NCM) causes severe numerical instabilities (note that we need NCM thanks to spin-orbit coupling);
with the GGA+NCM, we were not able to reach the level of accuracy needed to resolve the magnetic anisotropy energy, which could be as small as a few tens of $\muup$eV per transition metal atom.
The GGA+NCM, of which a widely adopted formulation was proposed in ref~\citenum{hobbsPRB2000}, has “both formal and numerical problems”~\cite{scalmaniJCTC2012}.  
The computational issue most likely emerges from small regions in space where spin density is close to zero.  
Here, if the local spin polarization changes slowly in space passing through zero, its direction is reverted.  
This reversal in the direction of the local spin polarization is not a big deal in reality because the magnitude of the local spin polarization is zero.  
One remarkable peculiarity of the GGA+NCM is that even if the spin density changes smoothly over such regions, the GGA+NCM functional recognizes it as a kink, the gradient of which is singular, thus causing severe numerical instabilities.

We examined how this serious issue had been dealt with in some public software packages.  
Surprisingly, currently there is no systematic solution for these instabilities based on justifiable theory but only ad hoc methods have been made without sound theoretical justification.  
In the case of the Elk code~\cite{elk}, for example, the kinks were smoothed out by preventing the magnitude of magnetization density from being smaller than a threshold.  
We would have applied this prescription to our code if it had been possible to keep this threshold sufficiently small while attaining a reasonable degree of convergence.  
We found, however, that the threshold had to be fairly large for convergence, so that it would affect the small magnetic anisotropies that we wanted to determine as accurately as possible.
In the case of VASP~\cite{kressePRB1996}, this inherent instability of GGA+NCM can be dealt with by using its own variation (whose formal and conceptual validity is not discussed) different from the original formulation of ref~\citenum{hobbsPRB2000}.
In passing we note that Quantum ESPRESSO did not make such ad hoc, unjustifiable compromises and hence it transparently shows the serious instability problem inherent in the GGA+NCM.

In fact, the numerical instability of the GGA+NCM is deeply related with its formal problem, as the scheme completely ignores the contribution to the exchange-correlation energy due to the gradient of magnetization density perpendicular to the local magnetization direction, while the exact exchange-correlation energy is certainly affected by that contribution~\cite{eichPRB2013}.  
There have been a few papers addressing this issue and proposing different formulations of the GGA+NCM~\cite{scalmaniJCTC2012, desmaraisJCP2019}, 
but none of them really complied with the conditions that should be satisfied by the exact exchange-correlation energy;
by the way, one of them has been retracted~\cite{desmaraisJCP2021}.
In contrast, the LDA is free from such issues.  
Therefore, we decided to use the LDA instead of the GGA to meet our strict criterion for convergence and to avoid the risk of being affected by those inherent problems of the GGA+NCM scheme.

\section{Calculation details: dipolar anisotropy}
\label{supp_sec:dip_aniso}
The contribution of the magnetic dipole-dipole interactions between localized magnetic moments to our magnetic model is given by the usual expression:
\begin{equation}
\label{supp_eqn:H_dip}
H_{\rm dip} =
\frac{1}{2} \sum_{i=1}^N \sum_{j=1, j\ne{}i}^N \left[ \frac{\mat{m}_i \cdot \mat{m}_j}{r_{ij}^3} - \frac{3(\mat{m}_i \cdot \mat{r}_{ij})(\mat{m}_j \cdot \mat{r}_{ij})}{r_{ij}^5} \right].
\end{equation}
The paramagnetic susceptibilities of MnPS$_3$ and NiPS$_3$ are well described by the spin-only magnetic moments of $S=2.5$ and $S=1$, respectively~\cite{joyPRB1992}.
This indicates that $\mat{m}_i$ for \mnps{} and \nips{} is given by $2\mu_{\rm B}\mat{S}_i$, where $\mu_{\rm B}$ is the Bohr magneton.
In the case of \feps{} ($S=2$), on the other hand, $\mat{m}_i$ is also contributed by the orbital polarizations.
The $z$ component of $\mat{m}_i$ is now given by $m_{iz} = \mu_{\rm B}(2S_{iz} + L_{iz})$, while the $x$ and $y$ components have no orbital contributions and are given by $m_{ix} = 2\mu_{\rm B}S_{ix}$ and $m_{iy} = 2\mu_{\rm B}S_{iy}$, respectively.
Because $H_{\rm dip}$ is negligibly small compared to the magnetic anisotropy arising from $LS$ coupling and a nonzero $L_{iz}$ ($=\pm1$) affects only a fraction ($\sim$25\%) of $m_{iz}$, whether the orbital contributions (the terms that include $L_{iz}$) of $H_{\rm dip}$ are exactly treated or not is of no importance at all in discussing the magnetic ordering of \feps{}.
In our calculations, $H_{\rm dip}$ for \feps{} was evaluated with $\mat{m}_i =  2\mu_{\rm B}\mat{S}_i$ so that the spin-only part of $H_{\rm dip}$ was taken into account, but this term can also be dropped in investigating the magnetic ordering in \feps{} (see the negligible difference between Figures~2e and h of the main manuscript).

Here, we further look into the effects of the dipolar anisotropy on the magnetic anisotropy of \tmps{} by calculating $H_{\rm dip}$ for the ground-state magnetic configurations of bulk \tmps{} (the configurations shown in Figures~\ref{fig:energy_mapping_pattern}c, g, and f for \mnps{}, \feps{}, and \nips{}, or, equivalently, in Figures~2a--c in the main manuscript, respectively) and examining the differences between the values of $H_{\rm dip}$ evaluated with imposing different collinear spin directions ($x$, $y$, and $z$) for $\mat{m}_i$'s.
We circumvented the issue of dealing with summing over infinitely many pairs of the magnetic sites ($i$ and $j$ in eq~\ref{supp_eqn:H_dip}) by neglecting the terms of which the distance between the spins $r_{ij}$ is longer than a certain cutoff $r_{\rm cut}$.
The convergence of the dipolar anisotropy with respect to the cutoff distance was tested for several values of $r_{\rm cut}$ ranging from 5\,\AA{} to 50\,\AA{}.

\begin{figure*}
\begin{center}
\includegraphics[width=0.8\textwidth]{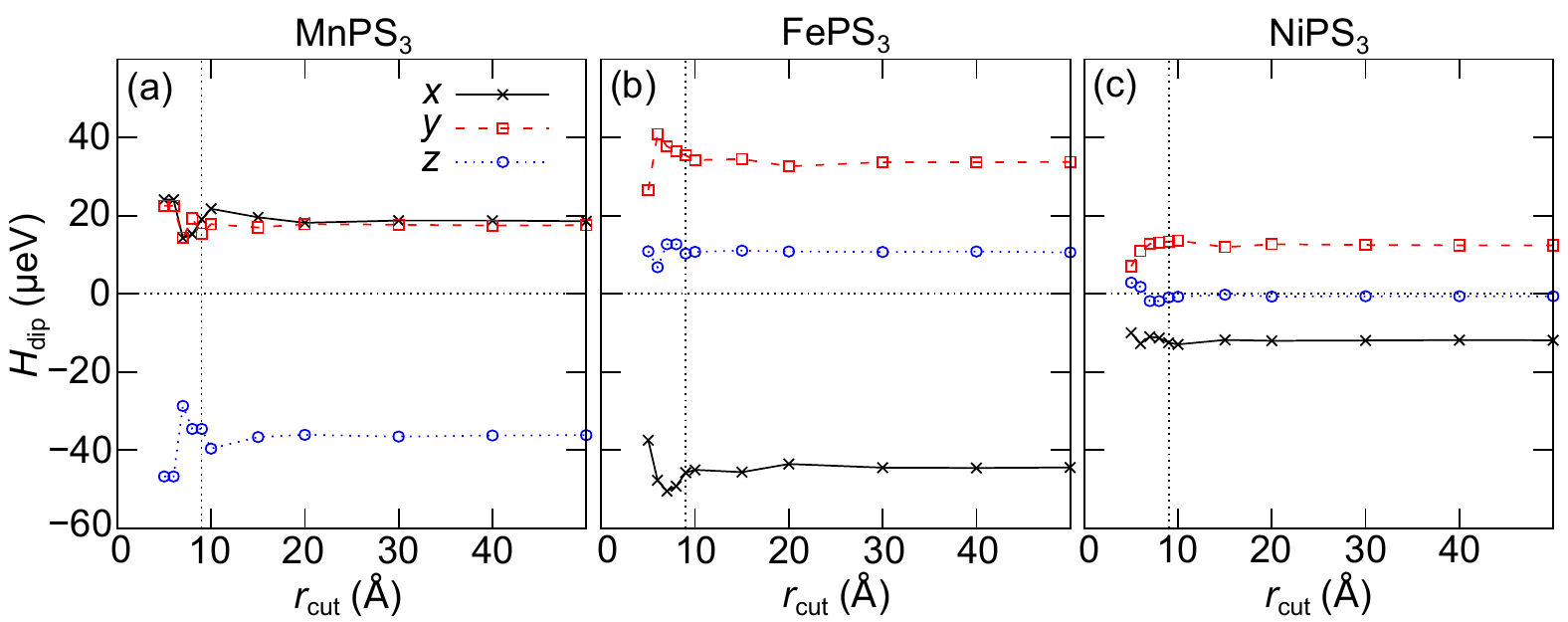}
\end{center}
\caption{
Convergence of the dipolar anisotropy, $H_{\rm dip}$ (eq~\ref{supp_eqn:H_dip}) per transition metal ion, with respect to the cutoff distance $r_{\rm cut}$.
The vertical dotted lines are placed at $r_{\rm cut}$ = 9\,\AA{}.
}
\label{fig:dip_energy}
\end{figure*}

As Figure~\ref{fig:dip_energy} shows,
the choice of $r_{\rm cut}$ = 9\,\AA{} yields reasonably converged results for the dipolar anisotropy of \tmps{} (see the vertical dotted lines). 
In the case of \mnps{} (Figure~\ref{fig:dip_energy}a), $H_{\rm dip}$ favors an easy-axis anisotropy along $z$.
In the case of \feps{} and \nips{} (Figures~\ref{fig:dip_energy}b and c), it gives rise to easy-axis anisotropies along the $x$ direction.
The magnitude of the dipolar anisotropy is about a few tens of $\muup$eV, ranging between 10\,$\muup$eV and 80\,$\muup$eV, per transition metal ion.

\section{Effects of \textit{LS} coupling on the magnetic anisotropy energy and some related numerical issues}
\label{supp_sec:ls_coupling}

\subsection{Effects of \textit{LS} coupling in FePS$_3$}
Given that $LS$ coupling dominantly contributes to the magnetic anisotropy of \feps{}, its (effective) functional form with respect to the orbital and spin polarizations can be assessed by looking at how the orbital polarization of an Fe ion and the total energy of the system change with the direction of the spin polarization.
Here, we present a simple analysis based on our DFT+$U$ calculations using fully-relativistic pseudopotentials that can justify the specific form of the $LS$ coupling term, $\sum_{i=1}^N \lambda L_{iz} S_{iz}$, used in our magnetic model (eq~\ref{supp_eqn:H_tot}).

We calculated the total energy of bilayer {\feps} and the orbital polarization of an Fe ion within the system for ferromagnetic configurations (Figure~\ref{fig:energy_mapping_pattern}a) of various collinear spin directions.
The choice of using the ferromagnetic ordering is just for convenience, because then all the transition metal ions have the same orbital and spin polarizations ($L_z = L_{iz}$ and $S_z = S_{iz}$ for all $i$).
We have checked that the results are essentially the same if we use the ground state magnetic configuration (Figure~\ref{fig:energy_mapping_pattern}g).

\begin{figure*}
\begin{center}
\includegraphics[width=0.6\textwidth]{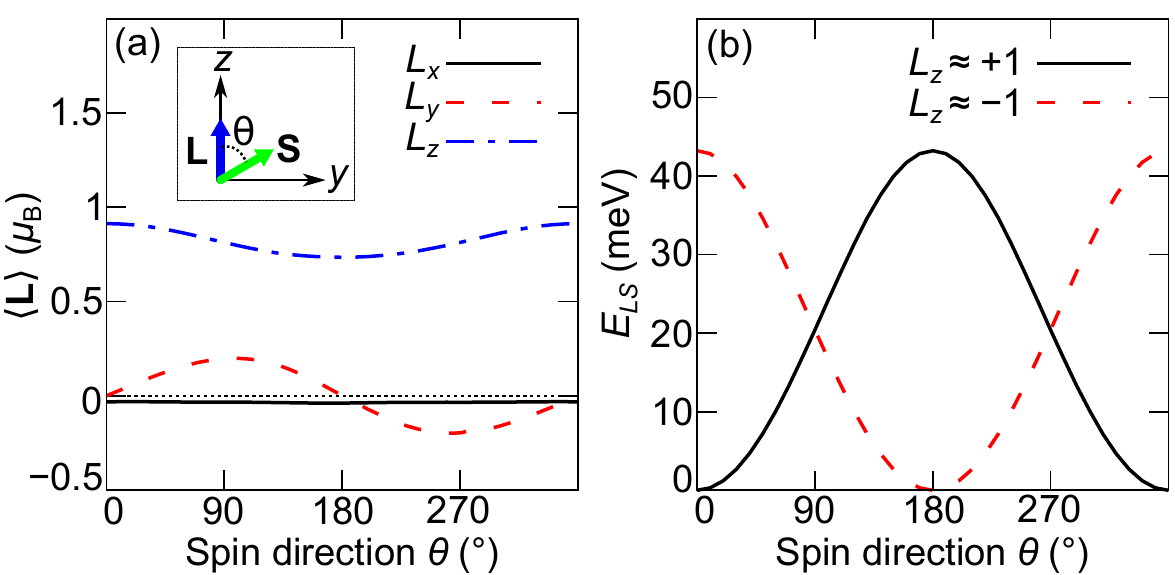}
\end{center}
\caption{
(a) Orbital polarization of an Fe ion in \feps{} as a function of the direction of its spin polarization.
Here, $\theta$ is the angle from $z$ (the out-of-plane direction of an \feps{} monolayer) of the spin polairzation direction.
(b) Magnetic anisotropy energy per Fe ion as a function of the spin polarization direction.
The solid black curve and the dashed red curve represent the case of an Fe ion with $L_z \approx 1$ and that of an Fe ion with $L_z \approx -1$, respectively.
The dash-dotted blue curve, which is slightly shifted downward for visual clarity, shows the lowest-energy at a given spin polarization direction. 
}
\label{fig:feps3_ls}
\end{figure*}

Figure~\ref{fig:feps3_ls}a shows the orbital polarization of an Fe ion as a function of the direction of its spin polarization.
The $z$ component of the orbital polarization is unquenched and is almost constant ($L_z \sim 1$) as the spin rotates fully (by 360$^\circ$ in the $yz$ plane as depicted) while the other components remain relatively small.
Figure~\ref{fig:feps3_ls}b displays the effect of this unquenched $L_z$ on the magnetic anisotropy.
It shows the total energy of the magnetic configurations with $L_z \approx 1$ and $L_z \approx -1$.
At $\theta$ = 90$^\circ$, where the spin lies in the $xy$ plane, there is a cusp in the lowest-energy curve (the dash-dotted blue curve in Figure~\ref{fig:feps3_ls}b), which is a direct consequence of the two-fold degeneracy arising from the orbital degrees of freedom ($L_z$ being either 1 or $-$1).
This $\left| \cos\theta \right|$ dependence on the collinear spin direction is differentiated from the usual $\cos^2\theta$ dependence  of the contributions arising from the other terms in eq~\ref{supp_eqn:H_tot}.
A similar analysis was done for the ground-state magnetic configuration (Figure~\ref{fig:energy_mapping_pattern}h or, equivalently, Figure~2b in the main manuscript) and the same $\left| \cos\theta \right|$ dependence was obtained.

We note that a delicate handling of the initial occupation matrix was required especially when we tried to calculate the total energy for the magnetic configurations whose local spins were aligned with the hard axes, the $x$ and $y$ directions.
And this numerical issue is closely related to the two-fold degeneracy shown at $\theta$ = 90$^\circ$ and 270$^\circ$ in Figure~\ref{fig:feps3_ls}b.
We had to initialize the occupation matrix with complex numbers so that the initial occupation matrix had unquenched orbital angular momentum (e.g. $L_z$ = $\pm1$),
and only through that way the correct, orbital-polarized ground states were obtained.
Since filling the initial occupation matrix with complex numbers is not the common practice for LDA+$U$ calculations,
it is likely that even experts in first-principles calculations might have missed this delicate total energy minimum.

\subsection{Failure of magnetic force theorem}
The influence of orbital degrees of freedom is not restricted to the sensitiveness to the initial condition on the occupation matrix.
What is more important is that it excludes the use of the magnetic force theorem (MFT)~\cite{liechtensteinJMMM1987} from consideration.
Since knowing how the MFT is working is important to understand why it should not be used in the case of \feps{}, 
we will shortly describe how the method is commonly implemented.  
First, a non-relativistic, self-consistent calculation is carried out to obtain well-converged collinear spin densities (two scalar quantities for up and down channels).  
Thus obtained spin densities have no preference for a specific direction, as they are only representing a collinear spin system.  
They are converted to a magnetization density (a vector quantity), and its collinear direction is then rotated toward a certain direction, usually the $x$, $y$ or $z$ direction.  
The rotated vector quantity is then used to initiate a fully-relativistic, non-self-consistent calculation.  
Such a one-shot calculation is repeated for other directions.  
Finally, the magnetic anisotropy energy is calculated from comparing the sum of the (Kohn-Sham) energy eigenvalues resulted from one of such non-self-consistent calculations with those for other directions.

The advantage of the MFT method is immediately noticed: 
because the method treats spin-orbit coupling as a perturbation, it circumvents all the intricacies that would be faced if a self-consistency with including the effects of spin-orbit coupling were to be achieved.  
The disadvantage by contrast is not that obvious.  
An important note here is that a non-relativistic calculation always yields spin densities without orbital degrees of freedom.  
(Note that the contribution to the orbital polarization of a Bloch state $\psi_{n\mathbf{k}}$ is cancelled by its complex conjugate---which is a symmetry operation of the Hamiltonian if spin-orbit coupling is absent---state with Bloch wavevector $-\mathbf{k}$.)  
While the MFT assumes implicitly that spin-orbit coupling is a small perturbation, the effects of spin-orbit coupling are significant in \feps{}, as they result in orbital polarizations, which are forbidden in non-relativistic treatments.  
The basic assumption of the MFT is thus no longer satisfied.  
In other words, the magnetic anisotropy of \feps{} should not be handled with the MFT.  

However, recent theoretical studies on \feps{} adopted the MFT and consequently reported much smaller (by almost two orders of magnitude) values for the magnetic anisotropy~\cite{nauman2M2021, olsenJPAP2021}, explaining why we had to resort to the computationally much trickier and heavier brute-force, total-energy method.  
Both our own MFT calculations on \feps{} and more involved total energy calculations without using a good (i.e., filled with properly designed complex elements) initial occupation matrix resulted in metastable states whose energy was quantitatively very similar to those reported in these two papers~\cite{nauman2M2021, olsenJPAP2021}.

\section{Isotropic interlayer exchange interactions}
Table~\ref{tab:inter_exch} shows the isotropic interlayer exchange interaction $K_n$ (= $1/3\,{\rm Tr}\,\mat{K}_n$) obtained from our calculations.
For all \tmps{}'s the second-nearest-neighbor isotropic interlayer exchange interactions ($K_{3,4}$) are relatively large compared to the other $K_n$'s.
Particularly in the case of \nips{}, $K_{3,4}$ are 1.64\,meV and as large as 30\,\% of $J_1^{(1\text{--}3)}$, indicating strong influence of the thickness on the magnetic properties of this compound.
It is noteworthy that superexchange paths for $K_n$'s need to go through at least two S ions, e.\,g.\,, like TM-S-S-TM, and that such pathways are similar in their total length for all $K_n$'s considered in this work.
This similarity partly explains how $K_{3,4}$ can be comparable to or even larger than the nearest-neighbor interlayer exchange interactions ($K_{1,2}$). 
Further investigation on the exchange mechanism behind $K_n$'s, which is beyond our current scope, would be helpful to address this issue, and we leave it for future studies.

\begin{table}
\begin{center}
\begin{tabular}{p{3em}*{8}{>{\raggedleft\arraybackslash}p{2.2em}}}
\hline\hline
& $K_1$ & $K_2$ & $K_3$ & $K_4$ & $K_5$ & $K_6$ & $K_7$ & $K_8$ \\
\hline
\mnps{} & 0.01 & 0.01 & 0.03 & 0.03 & 0.01 & 0.01 & 0.01 & 0.02 \\
\feps{} & 0.04 & 0.04 & 0.11 & 0.12 & 0.03 & 0.03 & 0.03 & 0.05 \\
\nips{} & $-$0.05 & $-$0.08 & 1.64 & 1.64 & 0.47 & 0.36 & 0.36 & 0.46 \\
\hline\hline
\end{tabular}
\caption{
Isotropic interlayer exchange interaction $K_n$ in units of meV.
}
\label{tab:inter_exch}
\end{center}
\end{table}

\section{Calculations details: classical Monte Carlo simulation}
In our classical Monte Carlo simulations, 
the Metropolis algorithm~\cite{metropolisJCP1953} and a variant of the Wolff cluster algorithm~\cite{wolffPRL1989}, modified for inclusion of the anisotropic exchange interactions~\cite{donoriodemeoPRB1992} and the single-ion anisotropy~\cite{leblancJPCM2013}, were used to update spins during simulations. 
For each measurement, one Metropolis sweep of the entire spin lattice and 10 Wolff cluster updates were done.
Typically, $10^4$--$10^5$ measurements were done for calculating thermodynamic quantities at a given temperature.
The initial 20--50\% of those measurements were discarded for thermalization.
In the case of bulk \tmps{}, periodic boundary conditions were imposed for all the crystallographic directions ($\mat{a}$, $\mat{b}'$, and $\mat{c}$ in Figure~\ref{fig:oblique_structure}).
We found that a superlattice containing $128 \times 128 \times 8$ unit cells including $128 \times 128 \times 8 \times 2 = 4096$ spins was sufficient to converge the results. 
In the case of few-layer \tmps{}, periodic boundary conditions were used for only the in-plane directions ($\mat{a}$ and $\mat{b}'$ in Figure~\ref{fig:oblique_structure}). 
A $128 \times 128 \times 1$ superlattice was employed for the few-layer cases. 

The magnetic susceptibility $\chi_\mu$ and the specific heat arising from the localized magnetic moments $c_v^{\rm mag}$ are defined as
\begin{equation}\label{supp_eqn:chi}
\chi_\mu = 
\frac{\langle M_\mu^2 \rangle - \langle M_\mu \rangle^2}{N T}
\end{equation}
and
\begin{equation}\label{supp_eqn:cv}
c_v^{\rm mag} = \frac{\langle H^2_\text{tot} \rangle - \langle H_\text{tot} \rangle^2}{N T^2},
\end{equation}
respectively, where $\mu$ denotes the Cartesian direction $x$, $y$, and $z$, $N$ is the number of magnetic atoms, $T$ is the temperature, and $M_\mu$ is the total magnetic moment. 
The expression $\left< O \right>$ denotes the thermal average of a quantity $O$, i.e. the average of $O$ with the Boltzmann weight factor.
In the case of \mnps{} and \nips{}, where the orbital degrees of freedom are suppressed ($L_{iz} = 0$) by the crystal field effects, $\left< O \right>$ is simply a weighted average of $O[\{ \mat{S}_i \}]$ over all spin configuration $\{ \mat{S}_i \}$'s.
It is straightforward to calculate this kind of weighted averages by using the importance sampling method, which is inherent in the process of updating spins by the algorithms we employed.
If additional degrees of freedom arising from the orbital polarization should be considered, as in the case of \feps{}, things become much more complicated because then $\left< O \right>$ involves averaging over all possible orbital configurations, $\{ L_{iz} \}$'s, as well as $\{ \mat{S}_i \}$'s.
In Section~\ref{supp_sec:orbital_suscept}, we discuss how to resolve this issue in practice.

\section{Orbital contributions to the magnetic susceptibility and specific heat}
\label{supp_sec:orbital_suscept}
We present a detailed description of how the effects of the orbital degrees of freedom on thermodynamic quantities were treated in our Monte Carlo simulations.
The basic idea is to integrate out orbital degrees of freedom in advance of averaging over spin degrees of freedom.
We first consider the partition function $Z$ of the anisotropic magnetic model $H_{\rm tot}$ (eq~\ref{supp_eqn:H_tot}):
\begin{equation}
Z = \sum_{\{\mat{S}_i\}} \sum_{\{L_{iz}\}} \left\{ \exp \left[ -\beta \left( H_{S {\rm only}} + \sum_{i=1}^N {\lambda L_{iz} S_{iz}} \right) \right] \right\}, \label{supp_eqn:Z}
\end{equation}
where $\beta$ is the inverse temperature $(k_\text{B} T)^{-1}$ and $H_{S {\rm only}}$ is the spin-only-dependent part of $H_{\rm tot}$ which reads
\begin{align}
H_{S {\rm only}} &= \frac{1}{2} \sum_{i=1}^{N} \sum_{n=1}^3 \sum_{a = 1}^{M_n} {\mat{S}_{i}^{\mathsf{T}}\,\mat{J}_{n}^{(a)}\,\mat{S}_{j(i,\mat{J}_n^{(a)})}} + \frac{1}{2} \sum_{i=1}^{N} \sum_{n=1}^8 {\mat{S}_{i}^{\mathsf{T}}\,\mat{K}_{n}\,\mat{S}_{j(i,\mat{K}_n)}} \nonumber \\
&+ \sum_{i=1}^{N} {\mat{S}_{i}^{\mathsf{T}}\,\mat{D}\,\mat{S}_{i}} + H_\text{dip}. \label{supp_eqn:H_S_only}
\end{align}
The summation over $\{L_{iz}\}$ in eq~\ref{supp_eqn:Z} can be done readily by exploiting that $L_{iz}$ can be either 1 or $-$1:
\begin{align}
Z &= \sum_{\{\mat{S}_i\}} \left\{ \exp (-\beta H_{S {\rm only}} ) \sum_{\{L_{iz}\}} \left[ \prod_{i=1}^N \exp \left(-\beta \lambda L_{iz} S_{iz} \right) \right] \right\} \nonumber \\
&= \sum_{\{\mat{S}_i\}} \left\{ \exp (-\beta H_{S {\rm only}} ) \prod_{i=1}^N \left[ \sum_{L_{iz} = \pm1} \exp \left(-\beta \lambda L_{iz} S_{iz} \right) \right] \right\} \nonumber \\
&= \sum_{\{\mat{S}_i\}} \left\{ \exp (-\beta H_{S {\rm only}} ) \prod_{i=1}^N 2\cosh\left(-\beta \lambda S_{iz} \right) \right\} \nonumber \\
&= \sum_{\{\mat{S}_i\}} \left\{ \exp \left[-\beta \left( H_{S {\rm only}} + \sum_{i=1}^N V_i (\beta, \lambda) \right) \right] \right\}, \end{align}
where $V_i(\beta,\lambda) = -\beta^{-1} \ln \left[2\cosh\left(-\beta\lambda S_{iz}\right)\right]$.
Here, $V_i(\beta,\lambda)$ can be thought of as an effective on-site potential that accounts for the effects of orbital degrees of freedom at a given temperature~$T$. 

The $z$ component of the total magnetic moment $\left< M_z \right> = \mu_{\rm B} \left< L_z + 2S_z \right>$, where $L_z = \sum_{i=1}^N L_{iz}$ and $S_z = \sum_{i=1}^N S_{iz}$.
The spin part can be written as
\begin{widetext}
\begin{align}
\left< S_z \right> &= Z^{-1} \sum_{\{ \mat{S}_i \}} \sum_{\{ L_{iz} \}} \left\{ S_{z} \exp \left[ -\beta \left( H_{S {\rm only}} + \sum_{i=1}^N \lambda L_{iz} S_{iz} \right) \right] \right\} \nonumber \\
&= Z^{-1} \sum_{\{ \mat{S}_i \}} \left\{ S_{z} \exp ( -\beta H_{S {\rm only}} ) \sum_{\{ L_{iz} \}} \left[ \prod_{i=1}^N \exp \left( -\beta \lambda L_{iz} S_{iz} \right) \right] \right\} \nonumber \\
&= Z^{-1} \sum_{\{ \mat{S}_i \}} \left[ S_{z} \exp ( -\beta H_{S {\rm only}} ) \prod_{i=1}^N 2\cosh \left( -\beta \lambda S_{iz} \right) \right] \nonumber \\
&= Z^{-1} \sum_{\{ \mat{S}_i \}} \left\{ S_{z} \exp \left[ -\beta \left( H_{S {\rm only}} + \sum_{i=1}^N V_i(\beta,\lambda) \right) \right] \right\} \nonumber \\
&= \left< S_z \right>_S. \label{supp_eqn:S_z}
\end{align}
Here, the notation $\left< O \right>_S$ was used to emphasize that the average was taken over spin configuration $\{ \mat{S}_i \}$'s only, with the weight factor $Z^{-1} \exp[-\beta (H_{S {\rm only}} + \sum_{i=1}^N V_i(\beta,\lambda) ]$.
The orbital part can be written as a sum over possible spin configurations as well, i.e.
\begin{align}
\left< L_z \right> &= Z^{-1} \sum_{\{ \mat{S}_i \}} \sum_{\{ L_{iz} \}} \left\{ L_{z} \exp\left[ -\beta \left( H_{S {\rm only}} + \sum_{i=1}^N \lambda L_{iz} S_{iz} \right) \right] \right\} \nonumber \\
&= Z^{-1} \sum_{\{ \mat{S}_i \}} \left\{ \exp(-\beta H_{S {\rm only}}) \sum_{\{ L_{iz} \}} \left[ L_{z}  \prod_{i=1}^N \exp(-\beta \lambda L_{iz} S_{iz}) \right] \right\} \nonumber \\
&= Z^{-1} \sum_{\{ \mat{S}_i \}} \left\{ \exp(-\beta H_{S {\rm only}}) \sum_{\{ L_{iz} \}} \left[ \sum_{j=1}^N L_{jz} \prod_{i=1}^N \exp(-\beta \lambda L_{iz} S_{iz}) \right] \right\} \nonumber \\
&= Z^{-1} \sum_{\{ \mat{S}_i \}} \left\{ \exp(-\beta H_{S {\rm only}}) \sum_{j=1}^N \left[ 2\sinh(-\beta \lambda S_{jz}) \prod_{i=1,i{\ne}j}^N 2\cosh(-\beta \lambda S_{iz}) \right] \right\} \nonumber \\
&= Z^{-1} \sum_{\{ \mat{S}_i \}} \left\{ \exp(-\beta H_{S {\rm only}}) \sum_{j=1}^N \left[ l_{jz} \prod_{i=1}^N 2\cosh(-\beta \lambda S_{iz}) \right] \right\} \nonumber \\
&= Z^{-1} \sum_{\{ \mat{S}_i \}} \left\{ \left( \sum_{j=1}^N l_{jz} \right) \exp\left[-\beta \left(H_{S {\rm only}} + \sum_{i=1}^N V_i(-\beta\lambda S_{iz})\right)\right] \right\} \nonumber \\
&= Z^{-1} \sum_{\{ \mat{S}_i \}} \left\{ l_z \exp\left[-\beta \left(H_{S {\rm only}} + \sum_{i=1}^N V_i(\beta,\lambda) \right) \right] \right\} \nonumber \\
&= \left< l_z \right>_S, \label{supp_eqn:L_z}
\end{align}
where $l_{iz} = \tanh(-\beta\lambda S_{iz})$ and $l_z = \sum_{i=1}^N l_{iz}$.

Some higher-order moments, such as $\langle L_z^2 \rangle$, $\langle S_z L_z \rangle$, and $\langle S_z^2 L_z^2 \rangle$, which are needed in calculating $\chi_z$ and $c_v^{\rm mag}$, are written as
\begin{align}
\langle L_z^2 \rangle &= Z^{-1} \sum_{\{ \mat{S}_i \}} \sum_{\{ L_{iz} \}} \left\{ L_{z}^2 \exp\left[ -\beta \left( H_{S {\rm only}} + \sum_{i=1}^N \lambda L_{iz} S_{iz} \right) \right] \right\} \nonumber \\
&= Z^{-1} \sum_{\{ \mat{S}_i \}} \left\{ \exp(-\beta H_{S {\rm only}}) \sum_{\{ L_{iz} \}} \left[ L_{z}^2  \prod_{i=1}^N \exp(-\beta \lambda L_{iz} S_{iz}) \right] \right\} \nonumber \\
&= Z^{-1} \sum_{\{ \mat{S}_i \}} \left\{ \exp(-\beta H_{S {\rm only}}) \sum_{\{ L_{iz} \}} \left[ \sum_{j=1}^N \sum_{k=1}^N L_{jz} L_{kz} \prod_{i=1}^N \exp(-\beta \lambda L_{iz} S_{iz}) \right] \right\} \nonumber \\
&= Z^{-1} \sum_{\{ \mat{S}_i \}} \left\{ \exp(-\beta H_{S {\rm only}}) \sum_{\{ L_{iz} \}} \left[\sum_{j=1}^N \left(1 + \sum_{k=1,k{\ne}j}^N L_{jz} L_{kz}\right) \prod_{i=1}^N \exp(-\beta \lambda L_{iz} S_{iz}) \right] \right\} \nonumber \\
&= Z^{-1} \sum_{\{ \mat{S}_i \}} \left\{ \exp(-\beta H_{S {\rm only}}) \sum_{j=1}^N \left[ \left( 1 + \sum_{k=1,k{\ne}j}^N l_{jz} l_{kz} \right) \prod_{i=1}^N 2\cosh(-\beta \lambda S_{iz}) \right] \right\} \nonumber \\
&= Z^{-1} \sum_{\{ \mat{S}_i \}} \left\{ \exp(-\beta H_{S {\rm only}}) \sum_{j=1}^N \left[ \left( 1 - l_{jz}^2 + \sum_{k=1}^N l_{jz} l_{kz} \right) \prod_{i=1}^N 2\cosh(-\beta \lambda S_{iz}) \right] \right\} \nonumber \\
&= Z^{-1} \sum_{\{ \mat{S}_i \}} \left\{ \left[ l_z^2 + \sum_{j=1}^N ( 1 - l_{jz}^2 ) \right] \exp\left[-\beta (H_{S {\rm only}} + \sum_{i=1}^N V_i(\beta,\lambda)\right] \right\} \nonumber \\
&= \left< l_z^2 + \sum_{i=1}^N ( 1 - l_{iz}^2 ) \right>_S, \label{supp_eqn:L_z^2}
\end{align}
\end{widetext}
\begin{align}
\langle S_z L_z\rangle = \left< S_z l_z\right>_S, \label{supp_eqn:S_zL_z}
\end{align}
and
\begin{align}
\langle S_z^2 L_z^2 \rangle = \left< S_z^2 \left\{l_z^2 + \sum_{i=1}^N ( 1 - l_{iz}^2 )\right\}\right>_S,
\label{supp_eqn:S_z^2L_z^2}
\end{align}
respectively.

Putting eqs~\ref{supp_eqn:S_z}--\ref{supp_eqn:S_z^2L_z^2} into eqs~\ref{supp_eqn:chi} and \ref{supp_eqn:cv},
we obtain the expressions for $\chi_z$ and $c_v^{\rm mat}$:
\begin{equation}
\chi_z = \mu_{\rm B}^2 \frac{
\langle
    (l_z + 2S_z)^2 
\rangle_S - 
\langle
    l_z + 2S_z 
\rangle_S^2 + 
\langle 
    \sum_{i=1}^N (1-l_{iz}^2) 
\rangle_S }{N T}.
\label{}
\end{equation}
and
\begin{equation}
c_v^\text{mag} = \frac{
    \langle
        H_{\rm tot, eff}^2 
    \rangle_S - 
    \langle
        H_{\rm tot, eff} 
    \rangle_S^2 + 
    \langle
        \lambda^2 \sum_{i=1}^N (1-l_{iz}^2) S_{iz}^2 
    \rangle_S }{N T^2},
\label{}
\end{equation}
respectively, where $H_{\rm tot, eff} = H_{S {\rm only}} + \sum_{i=1}^N \lambda l_{iz} S_{iz}$.

\section{Quantum fluctuation effects on the stability of the antiferromagnetic ground state of monolayer N\lowercase{i}PS$_3$}
The isotropic intralayer exchange interactions 
($J_n^{(a)}$, up to 28\,meV) are much larger in magnitude than other constituents of our magnetic Hamiltonian for \nips{}:
the anisotropic intralayer exchange interactions (the traceless part of $\mat{J}_n^{(a)}$, $\sim$0.07\,meV), the single-ion anisotropy ($\mathbf{D}$, $\sim$0.1\,meV), and the dipolar anisotropy energy ($\sim$0.025\,meV per transition metal ion).  
Therefore, we consider the quantum fluctuation effects in the $J_1$-$J_2$-$J_3$ Heisenberg model to gain insight into the possibility of a quantum spin liquid phase in monolayer \nips{}.

Because $S$=1 for the Ni$^{2+}$ ions ($3d^8$, high-spin configuration), the quantum effects of $S$=1 are of our particular interest.  
According to a study on the quantum ($S$=1/2 and $S$=1) $J_1$-$J_2$-$J_3$ model on the honeycomb lattice~\cite{merinoPRB2018}, the quantum spin liquid phase, which exists in a very small region in the parameter space in the case of $S$=1/2 (by the way, the sets of parameters for all three \tmps{} compounds are outside this small region), disappears in the case of $S$=1 due to the weakening of quantum fluctuations.  
A spin liquid phase driven by the frustration arising from competing $J_1$, $J_2$, and $J_3$ is therefore not likely to happen.

\begin{figure*}
\begin{center}
\includegraphics[width=0.6\textwidth]{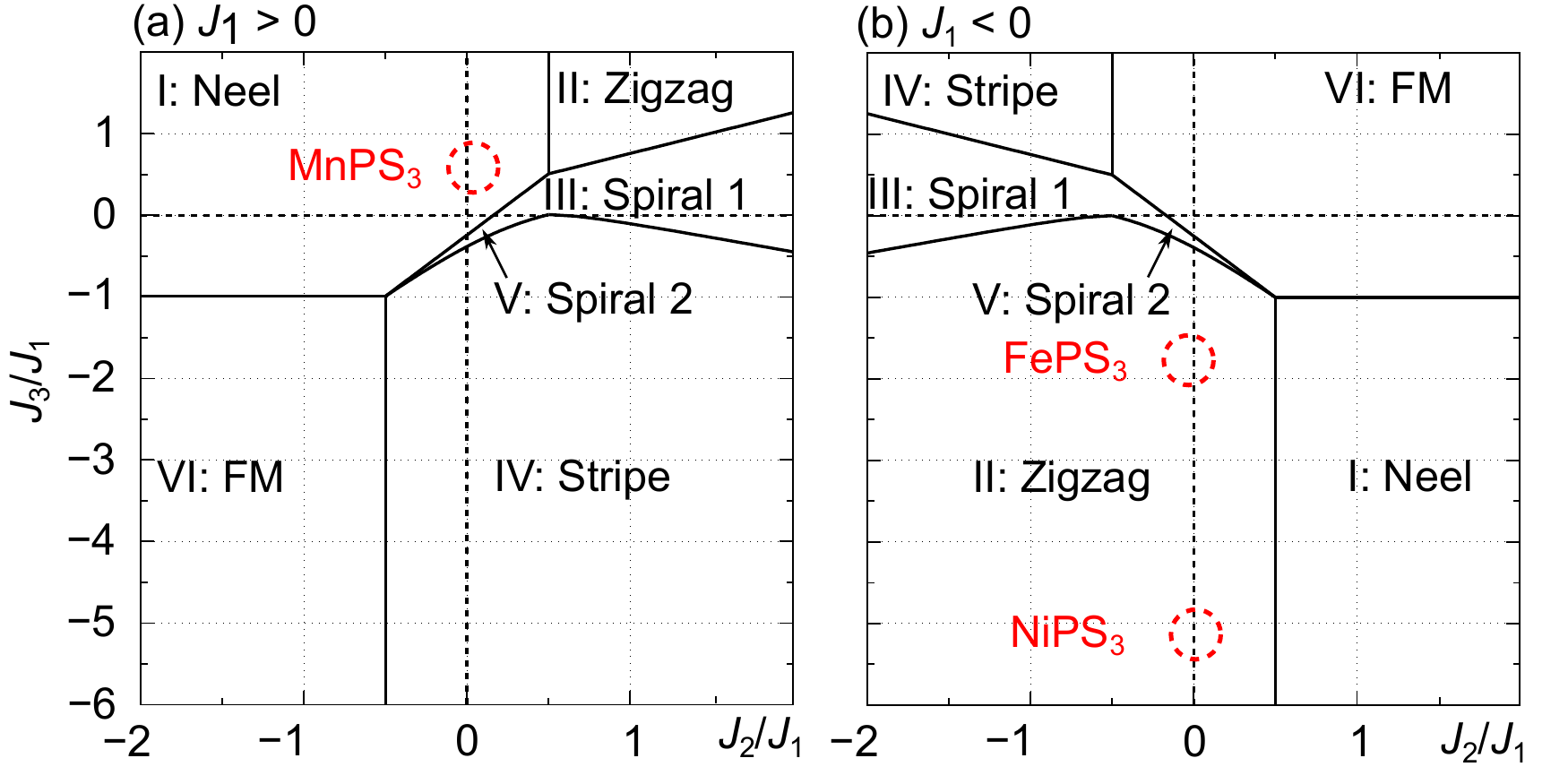}
\end{center}
\caption{
Phase diagrams of the classical $J_1$-$J_2$-$J_3$ model on the honeycomb lattice for (a) $J_1 > 0$ for MnPS$_3$ and (b) $J_1 < 0$ for FePS$_3$ and NiPS$_3$ (adapted from ref~\citenum{fouetEPJB2001}).
}
\label{fig:phase_diagram}
\end{figure*}

In addition to the suppression of the quantum spin liquid phase, the phase boundaries of the classical $J_1$-$J_2$-$J_3$ model on the honeycomb lattice (see Figure~\ref{fig:phase_diagram}) is only slightly changed when the quantum effects of $S$=1 are taken into account~\cite{merinoPRB2018}.  
Since our magnetic model for \nips{} is well-separated from any of those phase boundaries (see Figure~\ref{fig:phase_diagram}b),
the stability of the zigzag antiferromagnetic phase, which is the low temperature phase of monolayer \nips{} given by the classical $J_1$-$J_2$-$J_3$ model, would not be affected by such small quantum fluctuation effects.

For all these reasons, for monolayer \nips{}, the classical $J_1$-$J_2$-$J_3$ model provides almost the same answer as the quantum ($S$=1) $J_1$-$J_2$-$J_3$ model.  
Our use of classical Monte Carlo simulation can be justified in this regard.

Because the magnitude of the isotropic intralayer exchange interaction is much larger than that of the other contributions (the anisotropic intralayer exchange interactions, the single-ion anisotropy, and the dipolar anisotropy), we expect that our conclusion given above will not change when those minor contributions are fully considered within the quantum version ($S$=1) of our magnetic model for \nips{}.


%

\end{document}